\documentclass[12pt,a4paper]{article}
\usepackage{latexsym,graphicx,multirow}
\usepackage{float}
\usepackage{amssymb}
\usepackage{amscd}
\usepackage{amsthm}
\usepackage[left=2.5cm,top=2cm,right=2.5cm,bottom=2cm]{geometry}
\usepackage[hyperfootnotes=false]{hyperref}
\usepackage{epstopdf}
\usepackage{hyperref}
\usepackage{cite}
\usepackage{url}
\usepackage[utf8]{inputenc}
\usepackage{enumerate}
\usepackage{caption}
\usepackage{subcaption}
\usepackage{amsmath}
\usepackage[mathscr]{euscript}
\usepackage{framed}
\usepackage{xparse}
\usepackage{fancyvrb}
\usepackage{tikz}
\usetikzlibrary{patterns.meta}
\usepackage{multirow}
\usepackage{latexsym,graphicx,multirow}
\theoremstyle{plain}
\usepackage{longtable}
\usetikzlibrary{positioning,shapes}
\usepackage{algorithm}
\usepackage{algorithmic}
\newtheorem{theorem}{Theorem}[section]
\newtheorem{corollary}{Corrolary}[section]
\newtheorem{proposition}{Proposition}[section]
\newtheorem{remark}{Remark}[section]
\newtheorem{lemma}{Lemma}[section]
\newtheorem{definition}{Definition}[section]

\newcommand{\be}{\begin{equation}}
	\newcommand{\ee}{\end{equation}}
\newcommand{\bp}{\begin{proposition}}
	\newcommand{\ep}{\end{proposition}}
\newcommand{\ben}{\begin{equation*}}
	\newcommand{\een}{\end{equation*}}
\newcommand{\bd}{\begin{definition}}
	\newcommand{\ed}{\end{definition}}
\newcommand{\bl}{\begin{lemma}}
	\newcommand{\el}{\end{lemma}}
\newcommand{\bn}{\begin{notation}}
	\newcommand{\en}{\end{notation}}
\newcommand{\bcon}{\begin{construction}}
	\newcommand{\econ}{\end{construction}}
\newcommand{\bea}{\begin{eqnarray}}
	\newcommand{\eea}{\end{eqnarray}}
\newcommand{\bee}{\begin{eqnarray*}}
	\newcommand{\eee}{\end{eqnarray*}}
\newcommand{\bt}{\begin{theorem}}
	\newcommand{\et}{\end{theorem}}
\newcommand{\br}{\begin{remark}}
	\newcommand{\er}{\end{remark}}
\newcommand{\bo}{\begin{observation}}
	\newcommand{\eo}{\end{observation}}
\newcommand{\bex}{\begin{example}}
	\newcommand{\eex}{\end{example}}
\newcommand{\bc}{\begin{corollary}}
	\newcommand{\ec}{\end{corollary}}

\usepackage{etoolbox}
\NewDocumentCommand{\INTERVALINNARDS}{ m m }{
	#1 {,} #2
}

\makeatother
\NewDocumentCommand{\interval}{ s m >{\SplitArgument{1}{,}}m m o }{
	\IfBooleanTF{#1}{
		\left#2 \INTERVALINNARDS #3 \right#4
	}{
		\IfValueTF{#5}{
			#5{#2} \INTERVALINNARDS #3 #5{#4}
		}{
			#2 \INTERVALINNARDS #3 #4
		}
	}
}
\usepackage[symbol]{footmisc}
\renewcommand{\thefootnote}{\fnsymbol{footnote}}
\usepackage{cellspace} 
\begin{document}	
	\begin{center}
		\large{\bf{Estimation of the Coefficient of Variation of Weibull Distribution under Type-I Progressively Interval Censoring: A Simulation-based Approach}} \\
		\vspace{10mm}
		\normalsize{ Bankitdor M Nongrum$^{1}$, Adarsha Kumar Jena$^{2}$\footnote[1]{Corresponding author.} \let\thefootnote\relax\footnote{\textit{E-mail address:} jadarsha@gmail.com (Adarsha Kumar Jena)}
			 } \\
		\vspace{5mm}
		\normalsize{$^{1,2}$Department of Mathematics, National Institute of Technology Meghalaya, 793108, Meghalaya, India}\\
		\vspace{10mm}
	\end{center}
	
	\begin{abstract}
Measures of relative variability, such as the Pearson's coefficient of variation (CV$_p$), give much insight into the spread of lifetime distributions, like the Weibull distribution. The estimation of the Weibull CV$_p$ in modern statistics has traditionally been prioritized only when complete data is available. In this article, we estimate the Weibull CV$_p$ and its second-order alternative, denoted as CV$_k$, under type-I progressively interval censoring, which is a typical scenario in survival analysis and reliability theory. Point estimates are obtained using the methods of maximum likelihood, least squares, and the Bayesian approach with MCMC simulation. A nonlinear least squares method is proposed for estimating the CV$_p$ and CV$_k$. We also perform interval estimation of the CV$_p$ and CV$_k$ using the asymptotic confidence intervals, bootstrap intervals through the least squares estimates, and the highest posterior density intervals. A comprehensive Monte Carlo simulation study is carried out to understand and compare the performance of the estimators. The proposed least squares and the Bayesian methods produce better point estimates for the CV$_p$. The highest posterior density intervals outperform other interval estimates in many cases. The methodologies are also applied to a real dataset to demonstrate the performance of the estimators.
	\end{abstract}
	
	
		\section{Introduction}
		Measuring the variability of a statistical population is a significant aspect of modern statistical inference. When the variability of the population is measured relative to its mean, there is more concise information that gives much insight regarding its spread about the mean. The coefficient of variation (CV$_p$) is one such way to measure the relative variability of a distribution. Karl Pearson defined the CV$_p$ as the ratio of the standard deviation ($\sigma$) relative to the mean ($\mu \neq 0$). This measure can also be expressed as a percentage, i.e., CV$_p$ = $\sigma/\mu \times 100\%.$ The range of the CV$_p$ depends entirely on the distribution. It has an advantageous application for measuring variability as it is independent of the units of measurement. The higher the value of the CV$_p$, the greater the variability, and vice versa. The CV$_p$ has many applications in numerous fields. A few examples are given as follows: In medicine, it has been used as a key parameter for studying individual variation in susceptibility to SARS-CoV-2 disease (\cite{MGMG}). In chemistry, it is used for effective monitoring control in the chemical reactor process (\cite{TM}). In finance, the CV$_p$ is used to judge the relative risks of two stocks (\cite{EM}). It is also widely applicable in fields like survival analysis, see \cite{VS} for example, whereby its value could compare relative variability of datasets with complete different scales like mortality and cancer incidence. \\
		\indent There are times when the CV$_p$ is redefined in different types of settings as per the need of the situation. For instance, if one is interested only in knowing the absolute value of this variability, one would study the absolute coefficient of variation (\cite{KHB}). The inverse of the CV$_p$ (\cite{ANA}) is also of great interest to study, particularly for its application in image processing, where it is also called the Signal-to-Noise Ratio (SNR). Recently, there has been a newly developed variant of CV$_p$, called the second-order coefficient of variation \cite{TOK}), or Kvålseth's coefficient of variation (CV$_k$). This measure is defined as the ratio of the standard deviation ($\sigma$) to the square root of the second moment about the origin ($\mu_2'$), i.e., CV$_k = \sigma/\sqrt{\mu_2'}$. CV$_k$ is a non-dimensional value which lies between 0 and 1. It can also be expressed in terms of the CV$_p$ as: CV$_k^2 = \text{CV}_p^2/(1+\text{CV}_p^2)$. For our notation in this paper, we shall consider CV in general to represent either the CV$_p$ or CV$_k$. \\
		\indent The Weibull distribution is widely used in probability and statistical theory and other relevant practical domains. It is one of the most significant lifetime distributions and encompasses various applications. These applications span from survival analysis or reliability theory (\cite{AAEF}) to weather studies (\cite{MLo}), economics (\cite{MB}) and business administration (\cite{HA}), to name a few. 
		The shape parameter $\kappa$ and the rate parameter $\tau$ are non-negative reals, and we can write $T$ $\sim \text{Weibull}(\kappa,\tau)$ to represent a Weibull random variable $T$. Consider the two-parameter Weibull model whose cumulative distribution function (cdf) is given by:
		\begin{equation}
			\label{1.1} \tag{1.1}
			F(t;\kappa,\tau) = \begin{cases} 	1- \exp(-\tau t^\kappa) ,& t > 0 \\
			0, & t\leq 0\end{cases} 
		\end{equation}
		and the probability density function (pdf) is given by
		\begin{equation} \label{1.2}\tag{1.2}
			f(t;\kappa,\tau) =\begin{cases} \kappa\tau t^{\kappa - 1}  \exp(-\tau t^\kappa) ,& t > 0 \\
			0, & t\leq 0	\end{cases} 
		\end{equation}
		For this model, the respective CV$_p$ and CV$_k$ are:
		\begin{equation} \label{1.3} \tag{1.3}
			\text{CV}_p = \sqrt{\frac{\Gamma\big(1+\frac{2}{\kappa}\big)}{\{\Gamma\big(1+\frac{1}{\kappa}\big)\}^2}-1}~~\&~~\text{CV}_k = \sqrt{1-\frac{\{\Gamma\big(1+\frac{1}{\kappa}\big)\}^2}{\Gamma\big(1+\frac{2}{\kappa}\big)}}
		\end{equation}
		\indent Many researchers have conducted the statistical inferential study of the CV$_p$ of the Weibull distribution, but they have prioritized such inferences based on fully known, complete data. Many techniques have been developed for estimating the CV$_p$ in the past several years, where only a complete sample is considered. One example is the Bayesian method for estimating the CV$_p$ of a translated Weibull distribution proposed by \cite{WKP}. Their method uses a Markov Chain Monte Carlo (MCMC) simulation-based estimation strategy. The sequential sampling approach for estimating the CV$_p$ was created by \cite{BC} and used for the CV$_p$ of a Weibull distribution. Both \cite{II} and \cite{ML} have taken complete Weibull data into account and estimated the CV$_p$ using various interval estimation methods. Furthermore, for the Bayesian interval estimation of the CV$_p$, \cite{MLB} have implemented gamma and non-informative uniform priors. Although these previously published works have all used complete samples, this paper focuses on estimating the CV$_p$ and CV$_k$ using a censored data-set. In particular, the estimation of the CV$_p$ and CV$_k$, under a type-I progressively interval censoring scheme, is considered for our study.\\
		\indent The type-I progressively interval censoring scheme, introduced by \cite{RA}, is a form of censoring that progresses with time. The events of interest (sometimes called failures) are only observable within a fixed interval, although the time of occurrences of such events is unknown. This scheme is widely used when there are constraints over the number of times an event can be monitored or when its occurrence is only partially known to happen within a given time interval. Consider placing $n$ items at an initial time $t_0$ for some form of experimental monitoring, for instance, in a lifetime study. The items are considered for further monitoring at times $t_1, t_2,..., t_m$, which are all pre-specified. At each of the $t_i$'s, $i=1,2,...,m$, $W_i$ number of items are withdrawn from the experiment. These $W_i$'s may also be regarded as fixed percentages $p_1,p_2,...,p_m = 100\%$, of the remaining live subjects as the number of subjects remaining at each $t_i$ are all random. In addition, we also note the number $X_i$ of items that have failed, in the intervals $\interval({t_{i-1}, t_i}], i=1,...,m$. All the items that survive after the termination time $t_m$ are said to fail in the interval $(t_m,\infty)$.\\
		\indent It is known that $\sum_{i=1}^{m} (X_i + W_i) = n$. The general maximum likelihood function under this censoring scheme is given by:
		\begin{equation}
			\mathscr{L}(\boldsymbol{\theta}|\boldsymbol{t},\boldsymbol{X},\boldsymbol{W}) \propto \prod_{i=1}^{m} [F(t_i|\boldsymbol{ \theta}) - F(t_{i-1}|\boldsymbol{ \theta})]^{X_i} [1-F(t_i|\boldsymbol{ \theta})]^{W_i}\label{1.4}\tag{1.4}
		\end{equation}
		\indent Parameter estimation for the Weibull distribution under this censoring scheme has been done only on a small scale. For instance, \cite{HKTN} and \cite{CC} have presented different methods for parameter estimation, including the maximum likelihood estimation method. Bayesian inference has been done by \cite{YJL}, \cite{AK} and \cite{FA}. Further read can be found in \cite{CTL}, \cite{YU}. Some researchers have utilized this censoring scheme to obtain the estimators of the lifetime performance index (see \cite{SFW}) of products for a Weibull distribution. As no work has been done on the CV of the Weibull distribution under this censoring scheme, we present some point and interval estimation methods for this problem in this paper. \\
		\indent Henceforth, the paper will be organized as follows. In Section 2, we discuss the various methods of estimation. Within each subsection, we discuss both point and interval estimations. Section 3 is on data analysis, wherein we discuss the simulation study we conducted and compare the underlying results. In this section, we also study the methodologies using a real data set as an illustrative example. 
		\section{Methods of Estimation}
		\indent In each subsection of this section, we will discuss some of the techniques used for point estimation of parameters under type-I progressively interval censoring. Correspondingly, we will also provide the algorithms for determining the associated confidence intervals. Let $(t_i, X_i, W_i)$, $i=1,2,...,m$, be a type-I progressively interval censored data from a $\text{Weibull}(\kappa,\tau)$ distribution. Here, $W_i$ are the items progressively removed at each $t_i$ and $X_i$ are the items that have failed in the interval $\interval({t_{i-1}, t_i}], i=1,...,m$. Let $g(\kappa,\tau)$ be a function that stands for our parameters of interest: $\kappa$, $\tau$, CV$_p$ and CV$_k$.
		\subsection{Maximum Likelihood Estimation}
		At the very onset, we discuss the classical maximum likelihood estimation method. Following from equation (\ref{1.4}), the likelihood function for a Weibull distribution under the type-I progressively interval censoring is given below:
		\begin{equation} \label{2.1}\tag{2.1}
			\mathscr{L}(\kappa,\tau~|~\boldsymbol{t},\boldsymbol{X},\boldsymbol{W}) \propto \prod_{i=1}^{m}\big[\exp(\tau D_i)-1\big]^{X_i}\exp\big\{-\tau (X_i+W_i) t_i^\kappa\big\}   
		\end{equation}
		where, $D_i = t_i^\kappa - t_{i-1}^\kappa$. The log-likelihood function is then obtained and given as follows:
		\begin{equation} \label{2.2}\tag{2.2}
			\log\mathscr{L}(\kappa,\tau~|~\boldsymbol{t},\boldsymbol{X},\boldsymbol{W}) \propto \sum_{i=1}^{m}X_i\log\big[\exp(\tau D_i)-1\big] -\tau\sum_{i=1}^{m}(X_i + W_i)t_i^\kappa
		\end{equation}
		\indent Differentiating the log-likelihood function (\ref{2.2}) with respect to $\kappa$ and $\tau$, we respectively obtain the likelihood equations as follows. These equations are equated to zero, to obtain the MLEs of these parameters, denoted as $\hat{\kappa}$ and $\hat{\tau}$ respectively.
		\begin{align}
			\frac{\partial  \log\mathscr{L}}{\partial \kappa} = 0 \implies&  \sum_{i=1}^{m}\frac{X_i D_{i,\kappa}}{1-\exp(-\tau D_{i})} - \sum_{i=1}^{m}(X_i + W_i)t_i^\kappa \log t_i = 0 \label{2.3}\tag{2.3}\\
			\frac{\partial  \log\mathscr{L}}{\partial \tau} = 0 \implies& \sum_{i=1}^{m}\frac{X_i D_{i}}{1-\exp(-\tau D_{i})} - \sum_{i=1}^{m}(X_i + W_i)t_i^\kappa = 0\label{2.4}\tag{2.4}
		\end{align}
		\indent where, $D_{i,\kappa} = t_i^\kappa\log t_i - t_{i-1}^\kappa \log{t_{i-1}}$. It is not possible to analytically obtain the respective estimators in closed form. Therefore, these equations can be solved numerically by some well known root solving methods like the Newton-Raphson method. The above likelihood equations (\ref{2.3}-\ref{2.4}) are rewritten as $\ell_1(\kappa,\tau)=0$ and $\ell_2(\kappa,\tau)=0$ respectively. At the $r^\text{th}$ iteration, this method is given as:
		\begin{equation} \label{2.5}\tag{2.5}
			\begin{bmatrix}
				\kappa^{(r+1)} \\
				\tau^{(r+1)}
			\end{bmatrix} = \begin{bmatrix}
				\kappa^{(r)} \\
				\tau^{(r)}
			\end{bmatrix} - J^{-1}\begin{bmatrix}
				\ell_1(\kappa^{(r)},\tau^{(r)}) \\
				\ell_2(\kappa^{(r)},\tau^{(r)})
			\end{bmatrix}
		\end{equation}
		where,
		\begin{equation*} 
			J = \begin{bmatrix}
				\frac{\partial}{\partial\kappa} \ell_1(\kappa,\tau)      & \frac{\partial}{\partial\tau}\ell_1(\kappa,\tau)  \\
				\frac{\partial}{\partial\kappa}\ell_2(\kappa,\tau)  & \frac{\partial}{\partial\tau} \ell_2(\kappa,\tau)
			\end{bmatrix}
		\end{equation*}
		with,
		\begin{align*}
		\frac{\partial}{\partial\kappa}\ell_1(\kappa,\tau)       &= \sum_{i=1}^{m}\frac{X_i\big\{[1-\exp(-\tau D_i)]D_{i,\kappa\kappa} - \tau D_{i,\kappa}^2\exp(-\tau D_i)\big\}}{[1-\exp(-\tau D_i)]^2} - \sum_{i=1}^{m}(X_i + W_i)t_i^\kappa \log^2 t_i \\
			\frac{\partial}{\partial\tau}\ell_1(\kappa,\tau) &= -\sum_{i=1}^{m}\frac{X_i D_i D_{i,\kappa} \exp(-\tau D_i)}{[1-\exp(-\tau D_i)]^2} \\
			\frac{\partial}{\partial\kappa}\ell_2(\kappa,\tau)  &=  \sum_{i=1}^{m}\frac{X_i\big\{[1-\exp(-\tau D_i)]D_{i,\kappa} - \tau D_{i}D_{i,\kappa}\exp(-\tau D_i)\big\}}{[1-\exp(-\tau D_i)]^2} - \sum_{i=1}^{m}(X_i + W_i)t_i^\kappa \log t_i \\
			\frac{\partial}{\partial\tau} \ell_2(\kappa,\tau) &= -\sum_{i=1}^{m}\frac{X_i D_i^2  \exp(-\tau D_i)}{[1-\exp(-\tau D_i)]^2} 
		\end{align*}
		\indent By invariance property, the MLEs of the CV := CV$_p$ and CV := CV$_k$ are as follows:
		\begin{equation*}
			\widehat{\text{CV}}_p = \sqrt{\frac{\Gamma\big(\frac{2}{\hat{\kappa}} + 1\big)}{\{\Gamma\big(\frac{1}{\hat{\kappa}}+1\big)\}^2} - 1}~~~~\&~~~~
			\widehat{\text{CV}}_k = \sqrt{1 - \frac{\{\Gamma\big(\frac{1}{\hat{\kappa}}+1\big)\}^2}{\Gamma\big(\frac{2}{\hat{\kappa}} + 1\big)}} 
		\end{equation*}
	 Numerous researchers have pointed out that the Newton-Raphson converges slowly. \cite{HKTN} utilized the Expectation-Maximization (EM) algorithm as an improved method for deriving the MLEs of the Weibull parameters. However, \cite{CC} have presented an alternative maximum likelihood estimation algorithm that achieves a higher convergence rate than even the EM algorithm itself. This alternative approach utilizes mid-point approximation (see \cite{HKTN}) at the initial step and equivalent quantities to obtain the estimator for $\tau$. We will utilize this algorithm to obtain the MLE of $g(\kappa,\tau)$, which is given in Algorithm \ref{Al1} (see the next page). \\
\indent It is to remark that although this algorithm works effectively, it is essential that $\sum\limits_{i=1}^{m} X_i > 0$. This algorithm does not work if there are no failures in the experiment. \\
As the sample size increases, the MLE asymptotically converges weakly to the true parameter value. It is well known that under some regularity conditions, MLEs are asymptotically normally distributed, with the center of the normal distribution being the actual parameter value. Therefore, $\kappa \sim N(\kappa,\widehat{\text{var}}(\hat{\kappa}))$ and $\tau \sim N(\tau,\widehat{\text{var}}(\hat{\tau}))$. This property is useful for constructing confidence bounds for parameters and their functions, such as the CV. To do this, we first determine Fisher's information matrix at the very beginning to compute the information provided by the censored sample. Every $(i,j)^\text{th}$ entry in this matrix represents the expectation of the negative of the log-likelihood function's second partial derivatives, or $E[-\frac{\partial^2}{\partial \lambda_i\partial\lambda_j}\log\mathscr{L}]$, where $\lambda_i$ and $\lambda_j$ stand for the parameters $\kappa$ and $\tau$. However, the expectations are removed since evaluating them is complicated. The observed Fisher's information matrix $\hat{I}(\kappa,\tau)$ is defined as follows
		\begin{equation} \label{2.6}\tag{2.6}
			\hat{I}(\kappa,\tau) = \begin{bmatrix}
				-\frac{\partial^2}{\partial \kappa^2} \log\mathscr{L}       & 	-\frac{\partial^2}{\partial \kappa \partial \tau} \log\mathscr{L}   \\
				-\frac{\partial^2 }{\partial \tau \partial \kappa} \log\mathscr{L}      & 	-\frac{\partial^2}{\partial \tau^2} \log\mathscr{L}
			\end{bmatrix}_{\kappa = \hat{\kappa},~\tau=\hat{\tau}}
		\end{equation}
		where,
		\begin{align*}
			\frac{\partial^2\log\mathscr{L}}{\partial \kappa^2} &= \tau\sum_{i=1}^{m}X_i \frac{\{1-\exp(-\tau D_i)\}D_{i,\kappa \kappa} - \tau D_{i,\kappa}^2 \exp(-\tau D_i)}{[1-\exp(-\tau D_i)]^2} - \tau \sum_{i=1}^{m} (X_i + W_i) t_i^\kappa \log^2 t_i\\
			\frac{\partial^2\log\mathscr{L}}{\partial \kappa \partial \tau} &= \sum_{i=1}^{m}X_i \frac{\{1-\exp(-\tau D_i)\}D_{i,\kappa} - \tau D_i D_{i,\kappa} \exp(-\tau D_i)}{[1-\exp(-\tau D_i)]^2} - \sum_{i=1}^{m} (X_i + W_i) t_i^\kappa \log t_i \\
		\end{align*}
		\begin{align*}
			\frac{\partial^2\log\mathscr{L}}{ \partial \tau\partial \kappa} &= \frac{\partial^2\log\mathscr{L}}{\partial \kappa \partial \tau} \\
			\frac{\partial^2\log\mathscr{L}}{\partial \tau^2} &= -\sum_{i=1}^{m} X_i \frac{D_i^2 \exp(-\tau D_i)}{[1-\exp(-\tau D_i)]^2} 
		\end{align*}
				\begin{algorithm}[H]
			\caption{(Alternative Maximum Likelihood Estimation)}\label{Al1}
			\begin{algorithmic}[]
				\STATE 1. Specify tolerance value $\epsilon$ for estimating $\kappa$ and $\tau$.
				\STATE 2. Initialise $\kappa_p$ using the mid-point estimation method. \\
				\textbf{while} {$\epsilon > |\hat{\kappa}-\kappa_p|$} \textbf{do}
				\STATE \hspace*{1cm}3. Transfer intervals  $\interval({t_{i-1},t_i}],$ to  $\interval({t_{i-1}^\kappa, t_i^\kappa}]$, $i=1,...,m$ at $\kappa={\kappa}_p$.
				\STATE \hspace*{1cm}4. Initialise $\tau_p$ using the mid-point estimation method.\\
				\hspace*{1cm}\textbf{while} {$\epsilon > |\hat{\tau}-\tau_p|$} \textbf{do} 
				\STATE \hspace*{2cm} a. Calculate the equivalent failure time $t_{\interval({t_{i-1}^\kappa, t_i^\kappa}]}^*$
				\begin{equation} \label{2.7}\tag{2.7}
					t_{\interval({t_{i-1}^\kappa, t_i^\kappa}]}^* = \frac{1}{\hat{\tau}_p} + \frac{t_{i-1}^\kappa \exp(-\tau t_{i-1}^\kappa) - t_i^\kappa \exp(-\tau t_i^\kappa)}{\exp(-\tau t_{i-1}^\kappa) - \exp(-\tau t_{i}^\kappa)}
				\end{equation}
				\STATE \hspace*{2cm} b. Update $\tau$ as follows:
				\begin{equation} \label{2.8}\tag{2.8}
					\hat{\tau} = \frac{\sum\limits_{i=1}^{m} X_i}{\sum\limits_{i=1}^{m}X_i t_{\interval({t_{i-1}^\kappa, t_i^\kappa}]}^* + \sum\limits_{i=1}^{m}W_i t_i^\kappa}
				\end{equation}
				\STATE \hspace*{2cm} c. \textbf{if} {$|\hat{\tau}-\tau_p| < \epsilon$} \textbf{then} \\
				\STATE \hspace*{3cm} Go to Step 5. \\
				\hspace*{2cm} \textbf{else} \\
				\STATE \hspace*{3cm} $\tau_p = \hat{\tau}$. \\
				\hspace*{2cm} \textbf{end if} \\
				\hspace*{1cm} \textbf{end while}
				\STATE \hspace*{1cm}5. Find $\hat{\kappa} =  \arg \max\limits_{\kappa} \log \mathscr{L}(\kappa,\hat{\tau})$.
				\STATE \hspace*{1cm}6. \textbf{if}{$ |\hat{\kappa}-\kappa_p| < \epsilon$} \textbf{then}
				\STATE \hspace*{2cm} Go to Step 7. \\
				\hspace*{1cm} \textbf{else} \\
				\STATE \hspace*{2cm} $\kappa_p = \hat{\kappa}$. \\
				\hspace*{1cm} \textbf{end if}
				\STATE \hspace*{1cm}7. Determine $\widehat{\text{CV}}$. \\
				\textbf{end while}
			\end{algorithmic}
		\end{algorithm}
		\indent By inverting $\hat{I}(\kappa,\tau)$, we can obtain the symmetric asymptotic variance-covariance matrix $\hat{C}$ (see \cite{ACC}):
		\begin{equation}  \label{2.9}\tag{2.9}
			\hat{C} = \hat{I}^{-1} (\kappa,\tau)~\big|_{(\hat{\kappa},\hat{\tau})} =  \begin{bmatrix}
				\widehat{\text{var}}(\kappa)        & 	\text{cov}(\kappa,\tau)    \\
				\text{cov}(\tau,\kappa)       & 	\widehat{\text{var}}(\tau)
			\end{bmatrix}_{\kappa = \hat{\kappa},~\tau=\hat{\tau}}
		\end{equation}
		\indent To obtain the variance of $\widehat{\text{CV}}$, we make use of the asymptotic normal property of its MLE through the Delta method (\cite{WQM}). It can be easily shown that for the $\text{CV}$ with continuous first order partial derivative $\frac{\partial}{\partial \kappa}\text{CV}$ at $\kappa = \hat{\kappa}$,
		\begin{equation} \label{2.10}\tag{2.10}
			\sqrt{n}\{ \widehat{\text{CV}} -\text{CV}\} \xrightarrow{~~~~~L~~~~~} N(0,\hat{\sigma}_{\widehat{\text{CV}}}^2)
		\end{equation} 
		where, $\hat{\sigma}_{\widehat{\text{CV}}}^2 = [\frac{\partial}{\partial \kappa}{\text{CV}}]_{\kappa=\hat{\kappa}}^2 	\widehat{\text{var}}(\hat{\kappa})$. For $\text{CV} = $ CV$_p$,
		\begin{equation*} 
			\frac{\partial}{\partial\kappa}[\text{CV}_p] = \frac{\Gamma\bigg(1 + \frac{2}{\kappa}\bigg)}{\kappa^2 \text{CV}_p} \times \frac{\bigg[\Psi\bigg(1 + \frac{1}{\kappa}\bigg)-\Psi\bigg(1 + \frac{2}{\kappa}\bigg)\bigg]}{\Gamma^2\bigg(1 + \frac{1}{\kappa}\bigg)}
		\end{equation*}
		and for $\text{CV} = $ CV$_k$,
		\begin{equation*} 
			\frac{\partial}{\partial\kappa}[\text{CV}_k] = \frac{\Gamma^2(1+\frac{1}{\kappa})}{\kappa^2 \text{CV}_k}\times\bigg[\frac{\Psi\big(1+\frac{1}{\kappa}\big) - \Psi\big(1+\frac{2}{\kappa}\big)}{\Gamma\big(1+\frac{2}{\kappa}\big)}\bigg]
		\end{equation*}
		where $\Psi(x) = \frac{\Gamma^\prime(x)}{\Gamma(x)}$ denotes the digamma function. The standard $100(1-\beta)\%$ asymptotic confidence interval (ACI) for $\kappa$, $\tau$ and $\text{CV}$ are respectively expressed as 
		\begin{equation} \label{2.11}\tag{2.11}
		(\hat{\kappa} \pm  Z_{1-\frac{\beta}{2}} \sqrt{\widehat{\text{var}}(\hat{\kappa})}),~~(\hat{\tau} \pm Z_{1-\frac{\beta}{2}} \sqrt{\widehat{\text{var}}(\hat{\tau})})~~\text{and}~~( \widehat{\text{CV}} \pm Z_{1-\frac{\beta}{2}} \sqrt{\hat{\sigma}_{ \widehat{\text{CV}}}^2})
		\end{equation}
		where $Z_{1-\frac{\beta}{2}}$ denotes the upper $100(1-\frac{\beta}{2})^\text{th}$ percentile of the standard normal distribution. One disadvantage of this interval is that the lower bound can be negative, which is inappropriate since the parameters are strictly positive. This drawback can be easily corrected by correcting the lower bound of the ACI of $g(\kappa,\tau)$ ($=\kappa,$ $\tau$ or CV) as $\max(0,\hat{g}(\kappa,\tau) - Z_{1-\frac{\beta}{2}} \sqrt{\hat{\sigma}_{\hat{g}(\kappa,\tau)}^2})$. This method is, however, highly discouraged. An alternative way to overcome this is by using the log-transformation of the estimate $\hat{g}(\kappa,\tau)$ as follows:
		\begin{equation*} 
			\frac{\log \hat{g}(\kappa,\tau)-\log g(\kappa,\tau)}{\sqrt{\hat{\sigma}_{\log{\hat{g}(\kappa,\tau)}}^2}} \sim N(0,1)
		\end{equation*}
		where, $\hat{\sigma}_{\log{\hat{g}(\kappa,\tau)}}^2 = \frac{\hat{\sigma}_{\hat{g}(\kappa,\tau)}^2}{\hat{g}(\kappa,\tau)^2}$, for $g(\kappa,\tau)=\kappa$, $\tau$, and CV. The proof of this is similar by following the Delta method for $\log g(\kappa,\tau)$. Therefore, a $100(1-\beta)\%$ modified asymptotic confidence interval (MACI) for $\kappa$, $\tau$ and CV are respectively obtained as follows:
		\begin{align*}
			\bigg[\hat{\kappa}\exp\bigg\{-\frac{Z_{1-\frac{\beta}{2}} \sqrt{\widehat{\text{var}}(\hat{\kappa})}}{\hat{\kappa}} \bigg\},~& \hat{\kappa}\exp\bigg\{\frac{Z_{1-\frac{\beta}{2}} \sqrt{\widehat{\text{var}}(\hat{\kappa})}}{\hat{\kappa}} \bigg\}\bigg] \\
			\bigg[\hat{\tau}\exp\bigg\{-\frac{Z_{1-\frac{\beta}{2}} \sqrt{\widehat{\text{var}}(\hat{\tau})}}{\hat{\tau}} \bigg\},~& \hat{\tau}\exp\bigg\{\frac{Z_{1-\frac{\beta}{2}} \sqrt{\widehat{\text{var}}(\hat{\tau})}}{\hat{\tau}} \bigg\}\bigg] \\
			 \bigg[\widehat{\text{CV}}\exp{\bigg(-\frac{Z_{1-\frac{\beta}{2}}\sqrt{\hat{\sigma}_{{\widehat{\text{CV}}}}^2}}{\widehat{\text{CV}}}\bigg)},&~\widehat{\text{CV}}\exp{\bigg(\frac{Z_{1-\frac{\beta}{2}}\sqrt{\hat{\sigma}_{{\widehat{\text{CV}}}}^2}}{\widehat{\text{CV}}}\bigg)}\bigg]  \label{2.12}\tag{2.12}
		\end{align*}
		\indent The Algorithm \ref{Al2} (see the next page) is to be followed for obtaining the ACI and MACI for $g(\kappa,\tau)$.		
		\begin{algorithm}[t]
			\caption{(ACI and MACI)}\label{Al2}
			\begin{algorithmic}
				\STATE 1. Use sample $(t_i, X_i, W_i)$ to determine $\hat{\kappa}$, $\hat{\tau}$ and $\widehat{\text{CV}}$.
				\STATE 2.  Determine $\hat{I}$ using the equation (\ref{2.6}) and find its inverse $\hat{C}$.
				\STATE 3.  Apply the Delta Method to determine ${\hat{\sigma}^2}_{\widehat{\text{CV}}}$ and $\hat{\sigma}_{\log{\widehat{\text{CV}}}}^2$.
				\STATE 4. Determine the $100(1-\frac{\beta}{2})^\text{th}$ percentile  $Z_{1-\frac{\beta}{2}}$, where $Z\sim N(0,1)$.
				\STATE 5. A $100(1-\beta)\%$ symmetrical ACI for  $g(\kappa,\tau)=\kappa,$ $\tau$ and CV is
				\begin{equation*}
					[\max(0,\hat{g}(\kappa,\tau) - Z_{1-\frac{\beta}{2}} \sqrt{\hat{\sigma}_{\hat{g}(\kappa,\tau)}^2}),\hat{g}(\kappa,\tau) + Z_{1-\frac{\beta}{2}} \sqrt{\hat{\sigma}_{\hat{g}(\kappa,\tau)}^2}]
				\end{equation*}
				while a $100(1-\beta)\%$ MACI for  $g(\kappa,\tau)=\kappa,$ $\tau$ and CV is
				\begin{equation*}
					\bigg[\hat{g}(\kappa,\tau)\exp{\bigg(-\frac{Z_{1-\frac{\beta}{2}}\sqrt{\hat{\sigma}_{{\hat{g}(\kappa,\tau)}}^2}}{\hat{g}(\kappa,\tau)}\bigg)},~~\hat{g}(\kappa,\tau)\exp{\bigg(\frac{Z_{1-\frac{\beta}{2}}\sqrt{\hat{\sigma}_{{\hat{g}(\kappa,\tau)}}^2}}{\hat{g}(\kappa,\tau)}\bigg)}\bigg]
				\end{equation*}
			\end{algorithmic}
		\end{algorithm}
		\subsection{Least Squares Estimation}
		\indent The least squares estimation method can be approached in various ways. Over the years, the least squares strategy of estimating the parameters for the Weibull distribution has gradually developed under different censoring schemes. For example, \cite{XJ} has proposed different point and interval based-estimation methods for reliability estimation under right censoring. For type-I progressively interval censoring, some authors (see \cite{HKTN}) have considered employing the Weibull probability plot to obtain graphical estimates of the parameters. However, in the subsequent subsection, we will update this approach to achieve the linear least squares estimators. Furthermore, we have also proposed a non-linear least squares approach weighted by the number of censored units.
		\begin{enumerate}[(i)]
			\item \textit{Linear Least Squares Estimation (LLSE)}: The linear least squares estimating method is a popular methodology in statistical inference. A simple linear regression model may be expressed as $\boldsymbol{y} = \boldsymbol{C\Theta}+\boldsymbol{\varepsilon}$, where it is assumed that $\boldsymbol{\varepsilon}$ is a normally distributed error term. For the Weibull model, $\boldsymbol{y} = \log[-\log\{1-F(\boldsymbol{t})\}]$, $\boldsymbol{\Theta}= [\log{\tau},~\kappa]^T$ and 
			\begin{equation*}
				\boldsymbol{C} = \begin{bmatrix}
					1       & \log t_1 \\
					1       & \log t_2 \\
					\vdots  & \vdots \\
					1       & \log t_m
				\end{bmatrix}.
			\end{equation*}
			The residual sum of squares is defined as:
			\begin{equation}\label{2.13}\tag{2.13}
				S_{SE} = (\hat{\boldsymbol{y}} - \boldsymbol{C\Theta})^T(\hat{\boldsymbol{y}} - \boldsymbol{C\Theta})
			\end{equation}
		 	where, $\hat{y}_i = \log[-\log\{1-\hat{F}_i\}]$, $i=1,2,...,m$, with $\hat{F}_i$ being the non-parametric estimator of $F(t_i)$. To obtain the linear least squares estimates of the parameters, we minimize $S_{SE}$ and we therefore obtain:
			\begin{align*}
				\boldsymbol{\Theta} &= (\boldsymbol{C}^T\boldsymbol{C})^{-1}\boldsymbol{C}^T\hat{\boldsymbol{y}} \\
				&= \begin{bmatrix}
				\frac{	\sum\limits_{i=1}^{m}y_i\sum\limits_{i=1}^{m}\log^2 t_i - \sum\limits_{i=1}^{m} \log t_i\sum\limits_{i=1}^{m}y_i\log t_i}{m\sum\limits_{i=1}^{m}\log^2 t_i - (\sum\limits_{i=1}^{m} \log t_i)^2}      \\
				\frac{	m\sum\limits_{i=1}^{m}y_i \log t_i - \sum\limits_{i=1}^{m}y_i\sum\limits_{i=1}^{m} \log t_i}{m\sum\limits_{i=1}^{m}\log^2 t_i - (\sum\limits_{i=1}^{m} \log t_i)^2}     \\
				\end{bmatrix} \label{2.14}\tag{2.14}
			\end{align*}
		    or equivalently,
			\begin{align*}
				\tilde{\kappa}_l &= \frac{	m\sum\limits_{i=1}^{m}\hat{y}_i \log t_i - \sum\limits_{i=1}^{m}\hat{y}_i\sum\limits_{i=1}^{m} \log t_i}{m\sum\limits_{i=1}^{m}\log^2 t_i - (\sum\limits_{i=1}^{m} \log t_i)^2}  \\
				\tilde{\tau}_l &= \exp\bigg\{\frac{	\sum\limits_{i=1}^{m}\hat{y}_i\sum\limits_{i=1}^{m}\log^2 t_i - \sum\limits_{i=1}^{m} \log t_i\sum\limits_{i=1}^{m}\hat{y}_i\log t_i}{m\sum\limits_{i=1}^{m}\log^2 t_i - (\sum\limits_{i=1}^{m} \log t_i)^2}  \bigg\} \label{2.15}\tag{2.15}
			\end{align*}
			\cite{HKTN} have proposed a variation of the Kaplan-Meier estimator (otherwise known as the product limit estimator) to estimate $F(t_i)$:
			\begin{equation*}
				\hat{F}_i = 1 - \prod_{j=1}^{i} \bigg[1-\frac{X_j}{n-\sum\limits_{k=0}^{j-1}(X_k + W_k)}\bigg]
			\end{equation*} 
			where, $i=1,2,...,m$. However, if for some $j$, $X_j = n-\sum\limits_{k=0}^{j-1} (X_k + W_k)$, then $\hat{F}_i = 1$. As a result, $\hat{y}_i$ is undefined. Furthermore, if $X_1=0$, then $\hat{F}_1=0$, thus implying that $\hat{y}_1$ is undefined. Therefore, we opt for another estimator of $F(t_i)$. Recently, \cite{HHQ} have provided new methods to obtaining $\hat{F}$. One of these methods is the moments approximation method, and as such this estimator is given as:
			\begin{equation}  \label{2.16}\tag{2.16}
				\hat{F}_i = 1 - \prod_{j=m-i+1}^{m} \bigg[\frac{\sum\limits_{k=m-j+2}^{m}X_k + \sum\limits_{k=m-j+1}^{m}W_k + i}{{\sum\limits_{k=m-j+1}^{m}(X_k + W_k) + i+1}} \bigg]~~\forall~i=1,2,...,m
			\end{equation}
			\item \textit{Non-Linear Least Squares Estimation (NLLSE)}: We have also proposed a non-linear strategy, which is the non-linear least squares estimation method. This part is the main contribution of this paper. When the correlation between the variables is non-linear, as when working with the cumulative distribution function $F$, this approach works well especially for estimating CV. The censored units ($X_i,W_i$) are added as weights in the model. This method involves the direct minimization of the difference $S_o$ between the theoretical model and the observed data at two monitoring times $t_{i-1}$ and $t_i$, weighted by the number of failures in this interval. An additional weighted error is also considered at $t_i$ where the $W_i$ items that survive are withdrawn from the experiment. 
			\begin{align} 
				S_o &= \sum_{i=1}^{m} X_i[\{F(t_i)-F(t_{i-1})\} - \{\hat{F}_i - \hat{F}_{i-1}\}]^2 + \sum_{i=1}^{m} W_i[S(t_i) - \hat{S}_i]^2  \label{2.17}\tag{2.17} \\
				&= \sum_{i=1}^{m}\{(X_i + W_i)[1 - \exp(-\tau t_i^\kappa) - \hat{F}_i]^2 - X_i[1 - \exp(-\tau t_{i-1}^\kappa)-\hat{F}_{i-1}]^2\}  \label{2.18}\tag{2.18}
			\end{align}  
			subject to $\kappa$, $\tau >0$. Here, $S(t_i) = 1 - F(t_i)$, $i=1,2,...,m$. This procedure is equivalent to solving the following equations for the non-linear least squares estimators $\tilde{\kappa}_{nl}$ and $\tilde{\tau}_{nl}$:
			\begin{align*}
				\sum_{i=1}^{m}\{(X_i + W_i)t_i^{\tilde{\kappa}_{nl}}[1 - \exp(-\tilde{\tau}_{nl} t_i^{\tilde{\kappa}_{nl}}) - \hat{F}_i] \exp(-\tilde{\tau}_{nl}t_i^{\tilde{\kappa}_{nl}})- X_i t_{i-1}^{\tilde{\kappa}_{nl}}[1 -  &\exp(-\tilde{\tau}_{nl} t_{i-1}^{\tilde{\kappa}_{nl}}) \nonumber\\
				  - \hat{F}_{i-1}]\exp(-\tilde{\tau}_{nl} t_{i-1}^{\tilde{\kappa}_{nl}})\}&= 0\\
				\sum_{i=1}^{m}\{(X_i + W_i)t_i^{\tilde{\kappa}_{nl}}\log t_i[1 - \exp(-\tilde{\tau}_{nl} t_i^{\tilde{\kappa}_{nl}}) - \hat{F}_i] \exp(-\tilde{\tau}_{nl}t_i^{\tilde{\kappa}_{nl}})- 	X_i &t_{i-1}^{\tilde{\kappa}_{nl}}\log t_{i-1}\nonumber\\
				-\hat{F}_{i-1}] \exp(-\tilde{\tau}_{nl} [1 - \exp(-\tilde {\tau}_{nl} t_{i-1}^{\tilde{\kappa}_{nl}})t_{i-1}^{\tilde{\kappa}_{nl}})\} &= 0   \label{2.19}\tag{2.19}
			\end{align*}
			One can also optimize the unconstrained model by taking $\kappa = e^{\xi_\kappa}$ and $\tau = e^{\xi_\tau}$. Thus, the minimization problem is then given as:
			\begin{equation}
				\min S_o = \sum_{i=1}^{m}\{(X_i + W_i)[1 - \exp(-{e^{\xi_\tau}} t_i^{e^{\xi_\kappa}}) - \hat{F}_i]^2 - X_i[1 - \exp(-{e^{\xi_\tau}} t_{i-1}^{e^{\xi_\kappa}})-\hat{F}_{i-1}]^2\} \label{2.20}\tag{2.20}
			\end{equation}
			On substitution of the estimate $\tilde{\kappa}=\tilde{\kappa}_l$ or $\tilde{\kappa}_{nl}$, we can obtain the least squares estimate of the CVs as:
			\begin{equation*}
				\widetilde{\text{CV}}_p = \sqrt{\frac{\Gamma\big(\frac{2}{\tilde{\kappa}} + 1\big)}{\{\Gamma\big(\frac{1}{\tilde{\kappa}}+1\big)\}^2} - 1}~~~~\&~~~~
				\widetilde{\text{CV}}_k = \sqrt{1 - \frac{\{\Gamma\big(\frac{1}{\tilde{\kappa}}+1\big)\}^2}{\Gamma\big(\frac{2}{\tilde{\kappa}} + 1\big)}}
			\end{equation*}
		\end{enumerate}
		\indent We turn to bootstrapping to construct confidence intervals based on both least squares methods. The bootstrap samples are generated by setting the least squares estimates as the bootstrap parameters. The percentile bootstrap interval (PBI), which was first introduced by \cite{BE}, can be obtained easily. In this classical method, the bootstrap samples are generated using the pre-obtained MLEs of the original Weibull data. The bootstrap parameter estimates are obtained for each bootstrap sample. Furthermore, we obtain the quantiles of the bootstrap distribution to obtain the percentile bootstrap interval estimate for the parameters. We denote PBI-L and PBI-NL as the PBIs based on the linear and non-linear least squares estimators, respectively. The Algorithm \ref{Al3} (see the next page) is to be followed to construct the PBIs.
		\begin{algorithm}[h]
			\caption{(PBI)}\label{Al3}
			\begin{algorithmic}
				\STATE 1. Determine $\tilde{\kappa}$ and $\hat{\tau}$ from the given sample $(t_i,X_i,W_i)$.
				\STATE 2. Set $b=1$ and the number of bootstrap samples $B$.
				\WHILE{$b \leq B$}
				\STATE 3. Generate the bootstrap sample $(t_i,X_i^b,W_i^b)$ from the distribution $\text{Weibull}(\tilde{\kappa},\tilde{\tau})$.
				\STATE 4. Determine the bootstrap estimates $\{\tilde{g}(\kappa,\tau)\}^*_b$ and set $b=b+1$.
				\ENDWHILE
				\STATE 5. Sort $\{\tilde{g}(\kappa,\tau)\}^*_b$, $b=1,2,...,B$ in ascending order as:
				\begin{equation*}
					\{\tilde{g}(\kappa,\tau)\}^*_{(1)} \leq \{\tilde{g}(\kappa,\tau)\}^*_{(2)} \leq ... \leq \{\tilde{g}(\kappa,\tau)\}^*_{(B)}
				\end{equation*}
				\STATE 7. A $100(1-\beta)\%$ PBI for $g(\kappa,\tau)$ is
				\begin{equation} \label{2.21}\tag{2.21}
					\bigg[\{\tilde{g}(\kappa,\tau)\}^*\bigg(\frac{\beta}{2}\bigg),~~ \{\tilde{g}(\kappa,\tau)\}^*\bigg(1-\frac{\beta}{2}\bigg)\bigg]
				\end{equation}
				where $\{\tilde{g}(\kappa,\tau)\}^*\big(\frac{\beta}{2}\big) =\{\tilde{g}(\kappa,\tau)\}^*_{\big(\frac{\beta B}{2}\big)},$ which is the lower ${\big(\frac{\beta B}{2}\big)}^{\text{th}}$ percentile of the distribution of $\{\tilde{g}(\kappa,\tau)\}^*$.
			\end{algorithmic}
		\end{algorithm}
		\subsection{Bayesian Estimation}
		\indent In this subsection, we use Bayesian analysis to determine the point estimators and the highest posterior density interval (HPDI) for the parameters. This method requires specifying prior knowledge about the parameters and updating it with the available data. The HPDI is the smallest credible interval that contains the largest posterior probability density of the $g(\kappa,\tau)$. Moreover, the HPDI covers a broader range of values for the parameter of interest with higher probability through the posterior density. For more details, refer to \cite{MHC}. \\
		\indent The authors \cite{BCA} and \cite{SRH} have argued that no specific prior distribution can be definitively regarded superior to others when it comes to the selection of parameter-priors. The standard approach for the priors of both the Weibull parameters is to consider a pair of proper independent gamma priors for each parameter, see \cite{YA} for example. This prior choice is appropriate in the situation where the data is censored, because the likelihood function only provides a limited amount of information. Thus, let $\kappa~\sim~\text{Gamma}(a_1,a_2)$ and $\tau~\sim~\text{Gamma}(b_1,b_2)$, where $a_1, a_2, b_1, b_2 > 0$ are the hyper-parameters. The joint prior density is thus given as:
		\begin{equation} \label{2.22}\tag{2.22}
			\pi(\kappa,\tau) \propto \kappa^{a_1-1} {\tau}^{b_1 - 1} \exp(-a_2\kappa -b_2\tau)
		\end{equation}
		\indent When no prior information is available, one can opt for a non-informative prior like the Jeffrey's prior:
		\begin{equation*}
			\pi(\kappa,\tau) \propto \frac{1}{\kappa \tau}
		\end{equation*}
		\indent This prior distribution is a special case of the conventional gamma prior (hyper-parameters are set to zero). The conditional posterior density is then determined as given below: 
		\begin{equation} \label{2.23}\tag{2.23}
			\bar{\pi}(\kappa,\tau~|~\boldsymbol{t},\boldsymbol{X},\boldsymbol{W}) \propto \kappa^{a_1-1} \tau^{b_1-1}\exp\bigg[\sum_{i=1}^{m}X_i \log\{\exp(\tau D_i)-1\} -\tau\sum_{i=1}^{m}(X_i + W_i)t_i^\kappa -a_2\kappa - b_2\tau\bigg]
		\end{equation}
		\indent Evidently, obtaining closed-form expressions for the marginal posterior density of $\kappa$ and $\tau$ is too complicated. Therefore, we employ the Markov Chain Monte Carlo (MCMC) simulation method to determine the Bayes estimator of the parameters, including the CV, and subsequently, the HPDI for the same. The estimators of $\kappa$, $\tau$, and CV are determined using the Metropolis-Hastings (MH) Algorithm, a widely used Markov Chain Monte Carlo (MCMC) method. In this method, the unknown parameters, considered random variables, are derived from a less complicated density function known as the proposal density function. The fundamental concept of this technique is to construct a Markov Chain with a stationary distribution corresponding to the target density, as specified in \cite{DL}. \\
		\indent There are numerous methods for selecting the proposal density. Notable algorithms include the Independent MH (IMH) and the Random Walk MH (RWMH) algorithm. As its name suggests, the IMH algorithm chooses a proposal independent of the current state. We propose a bivariate Gaussian proposal in this algorithm. Specifically, the proposal candidate is $(\kappa',\tau')~\sim~\phi(\kappa,\tau)=N((\kappa,\tau),\boldsymbol{\Sigma})$, where $\boldsymbol{\Sigma}$ is the covariance matrix. At each step, the algorithm explores the state space globally. However, this algorithm is known for lagging and can get stuck easily.	The alternative method, the Random Walk Metropolis-Hastings (RWMH) algorithm, is therefore used in this paper. Each new state proposal in this algorithm is formed by applying a perturbation to the previous state. For convenience, we will use a bivariate Gaussian proposal in this algorithm. Unlike the IMH algorithm, this approach conducts a more local state space exploration at each step. \cite{EFS} implemented these algorithms to obtain the point estimates of the two Weibull parameters in the presence of right-censored data. \\
		The RWMH algorithm is given in Algorithm \ref{Al8}.
		   \begin{algorithm}[H]
			\caption{(Random Walk Metropolis Hastings Algorithm)}\label{Al8}
			\begin{algorithmic}
				\STATE 1. Take $(\kappa_{mp},\tau_{mp})$ as the choices of $(\kappa^{(0)}, {\tau}^{(0)})$.
				\STATE 2. Set $d=1$, the number of MCMC samples $M$ and the burn-in period $M_b$.
				\WHILE{$d \leq M$} 
				\STATE 3. Generate $(\kappa',\tau') = (\kappa^{(d-1)},\tau^{(d-1)}) + \boldsymbol{\epsilon}$, where $\boldsymbol{\epsilon} \sim N(\boldsymbol{0}, \boldsymbol{\Sigma})$. 
				\STATE 4. Compute $A_d = \min\bigg[1,\frac{\mathscr{L}(\kappa',\tau'~|~\boldsymbol{t},\boldsymbol{X},\boldsymbol{W})\pi(\kappa',\tau')}{\mathscr{L}(\kappa^{(d-1)},\tau^{(d-1)}~|~\boldsymbol{t},\boldsymbol{X},\boldsymbol{W})\pi(\kappa^{(d-1)},\tau^{(d-1)})}\bigg]$. 
				\STATE 5. Draw $u\sim U[0,1]$. If $u<A_l$, accept the candidate $(\kappa',\tau')$ and set $(\kappa^{(d)},\tau^{(d)}) = (\kappa',\tau')$, otherwise set $(\kappa^{(d)},\tau^{(d)}) = (\kappa^{(d-1)},\tau^{(d-1)})$. 
				\STATE 6. Determine $\text{CV}^{(d)}$ and set $d=d+1$. 
				\ENDWHILE
				\STATE 7. Sort $\{g(\kappa,\tau)\}^{(d)}$, $d=1,2,...,M$ in increasing order:
				\begin{equation*}
					\{g(\kappa,\tau)\}_{(1)} \leq \{g(\kappa,\tau)\}_{(2)} \leq \hdots \leq \{g(\kappa,\tau)\}_{(M)}
				\end{equation*} 
				\STATE 8. Determine $\check{g}(\kappa,\tau)$.
				\STATE 9. The endpoints of the $100(1-\beta)\%$ HPDI of $\{g(\kappa,\tau)\}$ are $\{g(\kappa,\tau)\}_{(h)}$ and $\{g(\kappa,\tau)\}_{(h+M(1-\beta))}$, where $h$ can be determined from the fact that:
				\begin{equation*}
					\{g(\kappa,\tau)\}_{(h+M(1-\beta))}- \{g(\kappa,\tau)\}_{(h)} = \min_{1\leq h^* \leq M\beta} [\{g(\kappa,\tau)\}_{(h^*+M(1-\beta))}- \{g(\kappa,\tau)\}_{(h^*)}]
				\end{equation*}
			\end{algorithmic}
		\end{algorithm}
		Suppose the MCMC samples of $g(\kappa,\tau)$ are $\{g(\kappa,\tau)\}^{(d)}$, $d=1,2,...,M$. These estimators are calculated by minimizing the squared error loss function, (see \cite{YA}), and after a certain burn-in period $M_b$, this estimate is given by:
        \begin{equation} \label{2.24}\tag{2.24}
		\check{g}(\kappa,\tau) = \frac{1}{M-M_b} \sum\limits_{d=M_b + 1}^{M}\{g(\kappa,\tau)\}^{(d)}
        \end{equation}
    	\indent It is important to note that the covariance matrix $\boldsymbol{\Sigma}$ cannot be assigned arbitrarily. The selection must ensure that the acceptance rate is neither excessively high nor excessively low. One can tune the $\boldsymbol{\Sigma}$ and perform some pilot runs until the resultant acceptance rate falls between 25 and 40\%. 

	\section{Data Analysis and Comparative Studies}
	\subsection{Simulation Studies}
	\indent We have conducted a comprehensive Monte-Carlo simulation study to understand and compare the performances of the various estimators. We have used the R software to perform the computational works. We have considered the statistical estimation for $\kappa$, $\tau$, CV$_p$ and CV$_k$. We have fixed values for the parameters as $(\kappa,\tau)$ $=(0.750,0.052)$ and $(1.250,0.525)$. When $\kappa = 0.750$, the fixed population CV$_p$ is $\text{CV}_p=1.3528$ and $\text{CV}_k=0.8041$. When $\kappa = 1.25$, $\text{CV}_p=0.805$ while $\text{CV}_k=0.6270$. Sample sizes ($n=50,200$) are considered for all the estimation methods. We generate a type-I progressively interval censored sample at two types of experimental settings, both starting at time $t_0 = 0$. 
	
	For the first case, there are $m=4$ examination points at $\boldsymbol{t} = (1,2,3,4)$, and for the second case, there are $m=8$ examination points at $\boldsymbol{t} = (1,2,3,4,5,6,7,8)$. The algorithm given by \cite{RA} is utilized for generating the samples. Furthermore, we have considered the following three choices of type-I progressively interval censoring schemes, based completely on the proportions of units that survived after each monitoring time:
			\begin{table}[h]
				\centering
				\caption{Sampling Schemes}
				\label{tab:my-table-1}
				\begin{tabular}{ccc}
					Scheme & $m=4$                       & $m = 8$                                    \\ \hline
					I  & $(0,0,0,1)$        & $ (0,0,0,0,0,0,0,1)$                 \\ \hline
					II & $(0.5,0,0,1)$      & $ (0.5,0,0,0,0,0,0,1)$             \\ \hline
					III & $(0.5, 0.5, 0.5, 1)$ & $ (0.1,0.1,0.1,0.1,0.1,0.1,0.1,1)$ \\ \hline
				\end{tabular}
			\end{table} \\
	\indent The schemes have been selected and ordered accordingly to illustrate the various censoring methods for both scenarios, as shown in Table \ref{tab:my-table-1}. For instance, in both cases, Scheme-I signifies no withdrawals at each examination time until the last termination time $t_m$. Scheme-II implies the only, but also heavy withdrawal at the first examination time $t_1$, and all surviving items at $t_m$ are withdrawn. Scheme III displays uniformity in the withdrawals of the surviving items at each succeeding examination time. It is also worth mentioning that in this simulation study, we have only considered those samples, whence the experiment did not complete prematurely before $t_m$. 
	
	 We have conducted $L=1000$ replications to calculate the mean square error (MSE), coverage probability (CP) and average interval width (AW) of each estimator as appropriate.
	
		For both bootstrap interval estimations, a resampling of $B = 2000$ times is considered. For the Bayes estimators (including the HPDI), each replication of a sample is being used to compute $M=50,000$ iterations along with a burn-in period of $M_b=5000$. The hyperparameters are chosen so that the prior means equal the original setting parameters.\\ That is,
		\begin{equation*}
			\frac{a_1}{a_2} = \kappa ~~~~\&~~~~ \frac{b_2}{b_2} = \tau
		\end{equation*}
	    \begin{center}
		\footnotesize
		\begin{longtable}[H]{|c|c|cccc|cccc|}
			\caption{MSEs of the estimates for CV$_p$ and CV$_k$ under various settings with $\kappa=0.750$} \label{tab:my-table-2} \\
			\hline
			\multicolumn{10}{|c|}{$\tau= 0.052$} \\
			\hline
			\multirow{3}{*}{\begin{tabular}[c]{@{}c@{}}Sampling \\ Scheme\end{tabular}}
			& \multirow{3}{*}{Method}
			& \multicolumn{4}{c|}{CV$_p$}
			& \multicolumn{4}{c|}{CV$_k$}
			\\ \cline{3-10}
			& & \multicolumn{2}{c|}{$n=50$} & \multicolumn{2}{c|}{$n=200$}
			& \multicolumn{2}{c|}{$n=50$} & \multicolumn{2}{c|}{$n=200$}
			\\ \cline{3-10}
			& & \multicolumn{1}{c|}{$m=4$} & \multicolumn{1}{c|}{$m=8$} & \multicolumn{1}{c|}{$m=4$} & $m=8$
			& \multicolumn{1}{c|}{$m=4$} & \multicolumn{1}{c|}{$m=8$} & \multicolumn{1}{c|}{$m=4$} & $m=8$
			\\ \hline
			\endfirsthead
			
			\hline
			\multicolumn{10}{|l|}{\textit{(continued from previous page)}} \\
			\hline
			\endhead
			
			\hline
			\multicolumn{10}{|r|}{\textit{Continued on next page}} \\
			\hline
			\endfoot
			
			\hline
			\endlastfoot
			
			\multirow{4}{*}{Scheme-I}
			& MLE & \multicolumn{1}{c|}{1.9461} & \multicolumn{1}{c|}{0.8272} & \multicolumn{1}{c|}{0.1953} & 0.0793 & \multicolumn{1}{c|}{0.0156} & \multicolumn{1}{c|}{0.0127} & \multicolumn{1}{c|}{0.0054} & 0.0026 \\ \cline{2-10}
			& LLSE & \multicolumn{1}{c|}{\textbf{0.1965}} & \multicolumn{1}{c|}{\textbf{0.1260}} & \multicolumn{1}{c|}{0.1197} & 0.0545 & \multicolumn{1}{c|}{0.0103} & \multicolumn{1}{c|}{0.0101} & \multicolumn{1}{c|}{0.0052} & 0.0028 \\ \cline{2-10}
			& NLLSE & \multicolumn{1}{c|}{0.1991} & \multicolumn{1}{c|}{0.1503} & \multicolumn{1}{c|}{\textbf{0.1100}} & \textbf{0.0544} & \multicolumn{1}{c|}{0.0160} & \multicolumn{1}{c|}{0.0148} & \multicolumn{1}{c|}{0.0056} & 0.0030 \\ \cline{2-10}
			& Bayes & \multicolumn{1}{c|}{1.3019} & \multicolumn{1}{c|}{0.3473} & \multicolumn{1}{c|}{0.3309} & 0.0846 & \multicolumn{1}{c|}{\textbf{0.0082}} & \multicolumn{1}{c|}{\textbf{0.0047}} & \multicolumn{1}{c|}{\textbf{0.0045}} & \textbf{0.0020} \\ \hline
			\multirow{4}{*}{Scheme-II}                                                  & MLE                     & \multicolumn{1}{c|}{1.8680}          & \multicolumn{1}{c|}{0.9438}          & \multicolumn{1}{c|}{1.2047}          & 0.1146          & \multicolumn{1}{c|}{0.0223}          & \multicolumn{1}{c|}{0.0185}          & \multicolumn{1}{c|}{0.0076}          & 0.0037          \\ \cline{2-10} 
			& LLSE                    & \multicolumn{1}{c|}{0.2359}          & \multicolumn{1}{c|}{\textbf{0.1747}} & \multicolumn{1}{c|}{0.1639}          & \textbf{0.0687} & \multicolumn{1}{c|}{0.0276}          & \multicolumn{1}{c|}{0.0176}          & \multicolumn{1}{c|}{0.0076}          & 0.0048          \\ \cline{2-10} 
			& NLLSE                   & \multicolumn{1}{c|}{\textbf{0.2356}} & \multicolumn{1}{c|}{0.1935}          & \multicolumn{1}{c|}{\textbf{0.1513}} & 0.0698          & \multicolumn{1}{c|}{0.0272}          & \multicolumn{1}{c|}{0.0204}          & \multicolumn{1}{c|}{0.0075}          & 0.0050          \\ \cline{2-10} 
			& Bayes                   & \multicolumn{1}{c|}{0.6121}          & \multicolumn{1}{c|}{0.3597}          & \multicolumn{1}{c|}{0.5802}          & 0.1311          & \multicolumn{1}{c|}{\textbf{0.0046}} & \multicolumn{1}{c|}{\textbf{0.0068}} & \multicolumn{1}{c|}{\textbf{0.0044}} & \textbf{0.0028} \\ \hline
			\multirow{4}{*}{Scheme-III}                                                 & MLE                     & \multicolumn{1}{c|}{0.9672}          & \multicolumn{1}{c|}{0.9360}          & \multicolumn{1}{c|}{2.3330}          & 0.0896          & \multicolumn{1}{c|}{0.0311}          & \multicolumn{1}{c|}{0.0162}          & \multicolumn{1}{c|}{0.0110}          & 0.0032          \\ \cline{2-10} 
			& LLSE                    & \multicolumn{1}{c|}{\textbf{0.3651}} & \multicolumn{1}{c|}{\textbf{0.1480}} & \multicolumn{1}{c|}{\textbf{0.1569}} & \textbf{0.0605} & \multicolumn{1}{c|}{0.0483}          & \multicolumn{1}{c|}{0.0139}          & \multicolumn{1}{c|}{0.0146}          & 0.0039          \\ \cline{2-10} 
			& NLLSE                   & \multicolumn{1}{c|}{0.3802}          & \multicolumn{1}{c|}{0.2173}          & \multicolumn{1}{c|}{0.1790}          & 0.0679          & \multicolumn{1}{c|}{0.0508}          & \multicolumn{1}{c|}{0.0232}          & \multicolumn{1}{c|}{0.0178}          & 0.0049          \\ \cline{2-10} 
			& Bayes                   & \multicolumn{1}{c|}{0.5149}          & \multicolumn{1}{c|}{0.3494}          & \multicolumn{1}{c|}{0.7529}          & 0.0962          & \multicolumn{1}{c|}{\textbf{0.0045}} & \multicolumn{1}{c|}{\textbf{0.0056}} & \multicolumn{1}{c|}{\textbf{0.0050}} & \textbf{0.0024} \\ \hline
			
			\multicolumn{10}{|c|}{$\tau= 0.525$} \\
			\hline
			\multirow{3}{*}{\begin{tabular}[c]{@{}c@{}}Sampling \\ Scheme\end{tabular}} & \multirow{3}{*}{Method}
			& \multicolumn{4}{c|}{CV$_p$}
			& \multicolumn{4}{c|}{CV$_k$} \\ \cline{3-10}
			& & \multicolumn{2}{c|}{$n=50$} & \multicolumn{2}{c|}{$n=200$} & \multicolumn{2}{c|}{$n=50$} & \multicolumn{2}{c|}{$n=200$} \\ \cline{3-10}
			& & \multicolumn{1}{c|}{$m=4$} & \multicolumn{1}{c|}{$m=8$} & \multicolumn{1}{c|}{$m=4$} & $m=8$
			& \multicolumn{1}{c|}{$m=4$} & \multicolumn{1}{c|}{$m=8$} & \multicolumn{1}{c|}{$m=4$} & $m=8$ \\ \hline
			
			\multirow{4}{*}{Scheme-I}
			& MLE  & \multicolumn{1}{c|}{0.1438} & \multicolumn{1}{c|}{0.0722} & \multicolumn{1}{c|}{0.0274} & 0.0163 & \multicolumn{1}{c|}{0.0041} & \multicolumn{1}{c|}{0.0025} & \multicolumn{1}{c|}{0.0011} & 0.0007 \\ \cline{2-10}
			& LLSE & \multicolumn{1}{c|}{\textbf{0.0834}} & \multicolumn{1}{c|}{\textbf{0.0442}} & \multicolumn{1}{c|}{\textbf{0.0244}} & \textbf{0.0147} & \multicolumn{1}{c|}{0.0040} & \multicolumn{1}{c|}{0.0024} & \multicolumn{1}{c|}{0.0010} & 0.0007 \\ \cline{2-10}
			& NLLSE & \multicolumn{1}{c|}{0.0843} & \multicolumn{1}{c|}{0.0535} & \multicolumn{1}{c|}{0.0245} & 0.0190 & \multicolumn{1}{c|}{0.0046} & \multicolumn{1}{c|}{0.0033} & \multicolumn{1}{c|}{0.0011} & 0.0010 \\ \cline{2-10}
			& Bayes & \multicolumn{1}{c|}{0.1654} & \multicolumn{1}{c|}{0.0811} & \multicolumn{1}{c|}{0.0292} & 0.0168 & \multicolumn{1}{c|}{\textbf{0.0030}} & \multicolumn{1}{c|}{\textbf{0.0020}} & \multicolumn{1}{c|}{\textbf{0.0010}} & \textbf{0.0006} \\ \hline
			
			\multirow{4}{*}{Scheme-II}
			& MLE  & \multicolumn{1}{c|}{0.3191} & \multicolumn{1}{c|}{0.0984} & \multicolumn{1}{c|}{0.0400} & 0.0229 & \multicolumn{1}{c|}{0.0059} & \multicolumn{1}{c|}{0.0027} & \multicolumn{1}{c|}{0.0015} & 0.0009 \\ \cline{2-10}
			& LLSE & \multicolumn{1}{c|}{0.1149} & \multicolumn{1}{c|}{\textbf{0.0431}} & \multicolumn{1}{c|}{0.0338} & \textbf{0.0187} & \multicolumn{1}{c|}{0.0052} & \multicolumn{1}{c|}{\textbf{0.0021}} & \multicolumn{1}{c|}{0.0015} & \textbf{0.0008} \\ \cline{2-10}
			& NLLSE & \multicolumn{1}{c|}{\textbf{0.1045}} & \multicolumn{1}{c|}{0.0503} & \multicolumn{1}{c|}{\textbf{0.0332}} & 0.0226 & \multicolumn{1}{c|}{0.0055} & \multicolumn{1}{c|}{0.0032} & \multicolumn{1}{c|}{0.0016} & 0.0011 \\ \cline{2-10}
			& Bayes & \multicolumn{1}{c|}{0.3214} & \multicolumn{1}{c|}{0.1362} & \multicolumn{1}{c|}{0.0450} & 0.0250 & \multicolumn{1}{c|}{\textbf{0.0037}} & \multicolumn{1}{c|}{0.0021} & \multicolumn{1}{c|}{\textbf{0.0014}} & 0.0009 \\ \hline
			
			\multirow{4}{*}{Scheme-III}
			& MLE  & \multicolumn{1}{c|}{0.8289} & \multicolumn{1}{c|}{0.0867} & \multicolumn{1}{c|}{0.0750} & 0.0187 & \multicolumn{1}{c|}{0.0075} & \multicolumn{1}{c|}{0.0027} & \multicolumn{1}{c|}{0.0023} & 0.0008 \\ \cline{2-10}
			& LLSE & \multicolumn{1}{c|}{\textbf{0.1068}} & \multicolumn{1}{c|}{\textbf{0.0455}} & \multicolumn{1}{c|}{\textbf{0.0518}} & \textbf{0.0168} & \multicolumn{1}{c|}{0.0069} & \multicolumn{1}{c|}{0.0024} & \multicolumn{1}{c|}{0.0026} & 0.0008 \\ \cline{2-10}
			& NLLSE & \multicolumn{1}{c|}{0.1900} & \multicolumn{1}{c|}{0.0647} & \multicolumn{1}{c|}{0.0519} & 0.0214 & \multicolumn{1}{c|}{0.0064} & \multicolumn{1}{c|}{0.0038} & \multicolumn{1}{c|}{0.0020} & 0.0009 \\ \cline{2-10}
			& Bayes & \multicolumn{1}{c|}{0.5400} & \multicolumn{1}{c|}{0.0995} & \multicolumn{1}{c|}{0.0904} & 0.0195 & \multicolumn{1}{c|}{\textbf{0.0040}} & \multicolumn{1}{c|}{\textbf{0.0021}} & \multicolumn{1}{c|}{\textbf{0.0019}} & \textbf{0.0007} \\ 
		\end{longtable}
	\end{center}
	\vspace{-1.5cm}
	\indent As such, we have approximated the prior hyperparameters by choice as $(a_1, a_2) = (3, 4)$ and $(5, 4)$for $\kappa=0.750$ and $1.250$ respectively. Similarly $(b_1, b_2) = (0.5, 10)$ and $(1, 2)$ for $\tau = 0.052$ and $0.525$ respectively.
	The point estimates of the CVs obtained are compared on the basis of their MSEs.
	Whereas, the interval estimates attained are compared through the coverage probabilities (CPs) and their average widths (AWs). The various methods in this paper are applied, and the result is analyzed in the next subsection. 
	\subsection{Discussions on Simulation Results}
	\indent In this subsection, we analyze the results obtained from the simulation study. We have tabulated only the results for CV, but we will mention the performances of the estimators of the parameters as well. We shall discuss the point estimators first. The MSEs of the estimators are given in Tables \ref{tab:my-table-2}-\ref{tab:my-table-3}. Naturally, the MSEs of all estimators decrease when the sample is larger in size. Therefore, through the simulation study, one can infer the 
		\begin{center}
		\footnotesize
		\begin{longtable}[H]{|c|c|cccc|cccc|}
			\caption{MSEs of the estimates for CV$_p$ and CV$_k$ under various settings with $\kappa = 1.250$} \label{tab:my-table-3} \\
			\hline
			\multicolumn{10}{|c|}{$\tau= 0.052$} \\
			\hline
			\multirow{3}{*}{\begin{tabular}[c]{@{}c@{}}Sampling \\ Scheme\end{tabular}}
			& \multirow{3}{*}{Method}
			& \multicolumn{4}{c|}{CV$_p$}
			& \multicolumn{4}{c|}{CV$_k$}
			\\ \cline{3-10}
			& & \multicolumn{2}{c|}{$n=50$} & \multicolumn{2}{c|}{$n=200$}
			& \multicolumn{2}{c|}{$n=50$} & \multicolumn{2}{c|}{$n=200$}
			\\ \cline{3-10}
			& & \multicolumn{1}{c|}{$m=4$} & \multicolumn{1}{c|}{$m=8$} & \multicolumn{1}{c|}{$m=4$} & $m=8$
			& \multicolumn{1}{c|}{$m=4$} & \multicolumn{1}{c|}{$m=8$} & \multicolumn{1}{c|}{$m=4$} & $m=8$
			\\ \hline
			\endfirsthead
			
			\hline
			\multicolumn{10}{|l|}{\textit{(continued from previous page)}} \\
			\hline
			\endhead
			
			\hline
			\multicolumn{10}{|r|}{\textit{Continued on next page}} \\
			\hline
			\endfoot
			
			\hline
			\endlastfoot
			\multirow{4}{*}{Scheme-I}                                                   & MLE                     & \multicolumn{1}{c|}{0.0756}          & \multicolumn{1}{c|}{0.0233}          & \multicolumn{1}{c|}{0.0161}          & 0.0058          & \multicolumn{1}{c|}{0.0118}          & \multicolumn{1}{c|}{0.0050}          & \multicolumn{1}{c|}{0.0033}          & 0.0012          \\ \cline{2-10} 
			& LLSE                    & \multicolumn{1}{c|}{0.0462}          & \multicolumn{1}{c|}{0.0178}          & \multicolumn{1}{c|}{0.0165}          & 0.0063          & \multicolumn{1}{c|}{0.0081}          & \multicolumn{1}{c|}{0.0039}          & \multicolumn{1}{c|}{0.0034}          & 0.0014          \\ \cline{2-10} 
			& NLLSE                   & \multicolumn{1}{c|}{\textbf{0.0380}} & \multicolumn{1}{c|}{0.0219}          & \multicolumn{1}{c|}{0.0142}          & 0.0059          & \multicolumn{1}{c|}{0.0084}          & \multicolumn{1}{c|}{0.0061}          & \multicolumn{1}{c|}{0.0031}          & 0.0014          \\ \cline{2-10} 
			& Bayes                   & \multicolumn{1}{c|}{0.0383}          & \multicolumn{1}{c|}{\textbf{0.0159}} & \multicolumn{1}{c|}{\textbf{0.0138}} & \textbf{0.0054} & \multicolumn{1}{c|}{\textbf{0.0044}} & \multicolumn{1}{c|}{\textbf{0.0029}} & \multicolumn{1}{c|}{\textbf{0.0025}} & \textbf{0.0011} \\ \hline
			\multirow{4}{*}{Scheme-II}                                                  & MLE                     & \multicolumn{1}{c|}{0.3059}          & \multicolumn{1}{c|}{0.0338}          & \multicolumn{1}{c|}{0.0221}          & 0.0079          & \multicolumn{1}{c|}{0.0156}          & \multicolumn{1}{c|}{0.0068}          & \multicolumn{1}{c|}{0.0041}          & 0.0017          \\ \cline{2-10} 
			& LLSE                    & \multicolumn{1}{c|}{0.0396}          & \multicolumn{1}{c|}{\textbf{0.0176}} & \multicolumn{1}{c|}{0.0178}          & 0.0076          & \multicolumn{1}{c|}{0.0090}          & \multicolumn{1}{c|}{0.0041}          & \multicolumn{1}{c|}{0.0037}          & 0.0018          \\ \cline{2-10} 
			& NLLSE                   & \multicolumn{1}{c|}{\textbf{0.0354}} & \multicolumn{1}{c|}{0.0181}          & \multicolumn{1}{c|}{\textbf{0.0160}} & 0.0073          & \multicolumn{1}{c|}{0.0079}          & \multicolumn{1}{c|}{0.0046}          & \multicolumn{1}{c|}{0.0034}          & 0.0017          \\ \cline{2-10} 
			& Bayes                   & \multicolumn{1}{c|}{0.0660}          & \multicolumn{1}{c|}{0.0221}          & \multicolumn{1}{c|}{0.0193}          & \textbf{0.0071} & \multicolumn{1}{c|}{\textbf{0.0053}} & \multicolumn{1}{c|}{\textbf{0.0035}} & \multicolumn{1}{c|}{\textbf{0.0030}} & \textbf{0.0015} \\ \hline
			\multirow{4}{*}{Scheme-III}                                                 & MLE                     & \multicolumn{1}{c|}{0.3844}          & \multicolumn{1}{c|}{0.0298}          & \multicolumn{1}{c|}{0.0478}          & 0.0072          & \multicolumn{1}{c|}{0.0202}          & \multicolumn{1}{c|}{0.0061}          & \multicolumn{1}{c|}{0.0070}          & 0.0015          \\ \cline{2-10} 
			& LLSE                    & \multicolumn{1}{c|}{\textbf{0.0362}} & \multicolumn{1}{c|}{\textbf{0.0173}} & \multicolumn{1}{c|}{\textbf{0.0211}} & 0.0066          & \multicolumn{1}{c|}{0.0109}          & \multicolumn{1}{c|}{0.0039}          & \multicolumn{1}{c|}{0.0050}          & 0.0015          \\ \cline{2-10} 
			& NLLSE                   & \multicolumn{1}{c|}{0.0400}          & \multicolumn{1}{c|}{0.0189}          & \multicolumn{1}{c|}{0.0248}          & 0.0072          & \multicolumn{1}{c|}{0.0127}          & \multicolumn{1}{c|}{0.0053}          & \multicolumn{1}{c|}{0.0064}          & 0.0017          \\ \cline{2-10} 
			& Bayes                   & \multicolumn{1}{c|}{0.0573}          & \multicolumn{1}{c|}{0.0194}          & \multicolumn{1}{c|}{0.0375}          & \textbf{0.0065} & \multicolumn{1}{c|}{\textbf{0.0048}} & \multicolumn{1}{c|}{\textbf{0.0033}} & \multicolumn{1}{c|}{\textbf{0.0043}} & \textbf{0.0013} \\ \hline
			\multicolumn{10}{|c|}{$\tau= 0.525$} \\\hline
			\multirow{3}{*}{\begin{tabular}[c]{@{}c@{}}Sampling \\ Scheme\end{tabular}} & \multirow{3}{*}{Method}
			& \multicolumn{4}{c|}{CV$_p$}
			& \multicolumn{4}{c|}{CV$_k$} \\ \cline{3-10}
			& & \multicolumn{2}{c|}{$n=50$} & \multicolumn{2}{c|}{$n=200$} & \multicolumn{2}{c|}{$n=50$} & \multicolumn{2}{c|}{$n=200$} \\ \cline{3-10}
			& & \multicolumn{1}{c|}{$m=4$} & \multicolumn{1}{c|}{$m=8$} & \multicolumn{1}{c|}{$m=4$} & $m=8$
			& \multicolumn{1}{c|}{$m=4$} & \multicolumn{1}{c|}{$m=8$} & \multicolumn{1}{c|}{$m=4$} & $m=8$ \\ \hline
			
			\multirow{4}{*}{Scheme-I}                                                   & MLE                     & \multicolumn{1}{c|}{0.0174}          & \multicolumn{1}{c|}{0.0435}          & \multicolumn{1}{c|}{0.0039}          & 0.0055          & \multicolumn{1}{c|}{0.0033}          & \multicolumn{1}{c|}{0.0065}          & \multicolumn{1}{c|}{0.0008}          & \textbf{0.0010} \\ \cline{2-10} 
			& LLSE                    & \multicolumn{1}{c|}{\textbf{0.0144}} & \multicolumn{1}{c|}{0.0834}          & \multicolumn{1}{c|}{0.0037}          & 0.0239          & \multicolumn{1}{c|}{0.0027}          & \multicolumn{1}{c|}{0.0114}          & \multicolumn{1}{c|}{\textbf{0.0008}} & 0.0041          \\ \cline{2-10} 
			& NLLSE                   & \multicolumn{1}{c|}{0.0241}          & \multicolumn{1}{c|}{\textbf{0.0326}} & \multicolumn{1}{c|}{0.0066}          & 0.0089          & \multicolumn{1}{c|}{0.0045}          & \multicolumn{1}{c|}{\textbf{0.0050}} & \multicolumn{1}{c|}{0.0014}          & 0.0017          \\ \cline{2-10} 
			& Bayes                   & \multicolumn{1}{c|}{0.0157}          & \multicolumn{1}{c|}{0.0448}          & \multicolumn{1}{c|}{\textbf{0.0008}} & \textbf{0.0058} & \multicolumn{1}{c|}{\textbf{0.0025}} & \multicolumn{1}{c|}{0.0060}          & \multicolumn{1}{c|}{0.0043}          & 0.0011          \\ \hline
			\multirow{4}{*}{Scheme-II}                                                  & MLE                     & \multicolumn{1}{c|}{0.0253}          & \multicolumn{1}{c|}{0.1149}          & \multicolumn{1}{c|}{0.0051}          & \textbf{0.0125} & \multicolumn{1}{c|}{0.0039}          & \multicolumn{1}{c|}{0.0141}          & \multicolumn{1}{c|}{0.0011}          & \textbf{0.0022} \\ \cline{2-10} 
			& LLSE                    & \multicolumn{1}{c|}{\textbf{0.0222}} & \multicolumn{1}{c|}{0.1677}          & \multicolumn{1}{c|}{\textbf{0.0048}} & 0.0551          & \multicolumn{1}{c|}{0.0035}          & \multicolumn{1}{c|}{0.0194}          & \multicolumn{1}{c|}{\textbf{0.0010}} & 0.0085          \\ \cline{2-10} 
			& NLLSE                   & \multicolumn{1}{c|}{0.0321}          & \multicolumn{1}{c|}{\textbf{0.0969}} & \multicolumn{1}{c|}{0.0090}          & 0.0189          & \multicolumn{1}{c|}{0.0053}          & \multicolumn{1}{c|}{\textbf{0.0116}} & \multicolumn{1}{c|}{0.0019}          & 0.0032          \\ \cline{2-10} 
			& Bayes                   & \multicolumn{1}{c|}{0.0243}          & \multicolumn{1}{c|}{0.1160}          & \multicolumn{1}{c|}{0.0050}          & 0.0137          & \multicolumn{1}{c|}{\textbf{0.0030}} & \multicolumn{1}{c|}{0.0126}          & \multicolumn{1}{c|}{0.0010}          & 0.0023          \\ \hline
			\multirow{4}{*}{Scheme-III}                                                 & MLE                     & \multicolumn{1}{c|}{0.0852}          & \multicolumn{1}{c|}{0.0572}          & \multicolumn{1}{c|}{0.0083}          & \textbf{0.0067} & \multicolumn{1}{c|}{0.0093}          & \multicolumn{1}{c|}{0.0081}          & \multicolumn{1}{c|}{0.0016}          & \textbf{0.0013} \\ \cline{2-10} 
			& LLSE                    & \multicolumn{1}{c|}{\textbf{0.0518}} & \multicolumn{1}{c|}{0.1038}          & \multicolumn{1}{c|}{\textbf{0.0081}} & 0.0349          & \multicolumn{1}{c|}{0.0071}          & \multicolumn{1}{c|}{0.0135}          & \multicolumn{1}{c|}{\textbf{0.0015}} & 0.0058          \\ \cline{2-10} 
			& NLLSE                   & \multicolumn{1}{c|}{0.0808}          & \multicolumn{1}{c|}{\textbf{0.0454}} & \multicolumn{1}{c|}{0.0135}          & 0.0085          & \multicolumn{1}{c|}{0.0096}          & \multicolumn{1}{c|}{\textbf{0.0064}} & \multicolumn{1}{c|}{0.0025}          & 0.0017          \\ \cline{2-10} 
			& Bayes                   & \multicolumn{1}{c|}{0.0766}          & \multicolumn{1}{c|}{0.0583}          & \multicolumn{1}{c|}{0.0086}          & 0.0072          & \multicolumn{1}{c|}{\textbf{0.0069}} & \multicolumn{1}{c|}{0.0074}          & \multicolumn{1}{c|}{0.0015}          & 0.0013      
		\end{longtable}
	\end{center}
	\vspace{-1.5cm}
		consistent nature of the estimators. Furthermore, the comparisons are made through various factors:
		\begin{enumerate}
			\item In terms of the sampling scheme plan, Scheme-I consistently delivers estimates with the lowest MSEs.
			\item In terms of the number of examination points, or rather the length of the experiment, it is observed that the longer ($m=8$) the experiment, the better the performance of the estimators of $\kappa$, CV$_p$, and CV$_k$. However, a similar trend for $\tau$ is only observed for its smaller setting value.
			
			\item Let $\tau$ be fixed. Regarding setting values of $\kappa$, the estimators when $\kappa=1.25$ generally perform better than when $\kappa=0.75$. There are a few exceptions. This scenario for the case of the MSEs of the $\kappa$ estimates can be seen only when $m=8$. However, the converse happens for $m=4$. Another exception is for the CV$_k$, as its estimates when $\tau=0.525$ have lower MSEs than when $\tau=0.052$. This result explains that
				\begin{center}
				\footnotesize
				\begin{longtable}[H]{|c|c|cccc|cccc|}
					\caption{Coverage Probabilities of the interval estimates for 95\% confidence of CV$_p$ and CV$_k$ under various settings with $\kappa = 0.750$} \label{tab:my-table-4} \\
					\hline
					\multicolumn{10}{|c|}{$\tau= 0.052$} \\
					\hline
					\multirow{3}{*}{\begin{tabular}[c]{@{}c@{}}Sampling \\ Scheme\end{tabular}}
					& \multirow{3}{*}{Method}
					& \multicolumn{4}{c|}{CV$_p$}
					& \multicolumn{4}{c|}{CV$_k$}
					\\ \cline{3-10}
					& & \multicolumn{2}{c|}{$n=50$} & \multicolumn{2}{c|}{$n=200$}
					& \multicolumn{2}{c|}{$n=50$} & \multicolumn{2}{c|}{$n=200$}
					\\ \cline{3-10}
					& & \multicolumn{1}{c|}{$m=4$} & \multicolumn{1}{c|}{$m=8$} & \multicolumn{1}{c|}{$m=4$} & $m=8$
					& \multicolumn{1}{c|}{$m=4$} & \multicolumn{1}{c|}{$m=8$} & \multicolumn{1}{c|}{$m=4$} & $m=8$
					\\ \hline
					\endfirsthead
					
					\hline
					\multicolumn{10}{|l|}{\textit{(continued from previous page)}} \\
					\hline
					\endhead
					
					\hline
					\multicolumn{10}{|r|}{\textit{Continued on next page}} \\
					\hline
					\endfoot
					
					\hline
					\endlastfoot
					
					\multirow{4}{*}{Scheme-I}                                                   & MACI                    & \multicolumn{1}{c|}{0.967}          & \multicolumn{1}{c|}{0.93}           & \multicolumn{1}{c|}{0.951}          & 0.959          & \multicolumn{1}{c|}{0.932}          & \multicolumn{1}{c|}{0.928}          & \multicolumn{1}{c|}{\textbf{0.945}} & 0.936          \\ \cline{2-10} 
					& PBI-L                   & \multicolumn{1}{c|}{0.947}          & \multicolumn{1}{c|}{0.73}           & \multicolumn{1}{c|}{0.932}          & 0.906          & \multicolumn{1}{c|}{0.947}          & \multicolumn{1}{c|}{0.73}           & \multicolumn{1}{c|}{0.932}          & 0.906          \\ \cline{2-10} 
					& PBI-NL                  & \multicolumn{1}{c|}{0.81}           & \multicolumn{1}{c|}{0.565}          & \multicolumn{1}{c|}{0.903}          & 0.873          & \multicolumn{1}{c|}{0.81}           & \multicolumn{1}{c|}{0.565}          & \multicolumn{1}{c|}{0.903}          & 0.873          \\ \cline{2-10} 
					& HPDI                    & \multicolumn{1}{c|}{\textbf{0.987}} & \multicolumn{1}{c|}{\textbf{0.976}} & \multicolumn{1}{c|}{\textbf{0.955}} & \textbf{0.968} & \multicolumn{1}{c|}{\textbf{0.974}} & \multicolumn{1}{c|}{\textbf{0.964}} & \multicolumn{1}{c|}{0.938}          & \textbf{0.946} \\ \hline
					\multirow{4}{*}{Scheme-II}                                                  & MACI                    & \multicolumn{1}{c|}{0.928}          & \multicolumn{1}{c|}{0.886}          & \multicolumn{1}{c|}{0.952}          & 0.948          & \multicolumn{1}{c|}{0.901}          & \multicolumn{1}{c|}{0.93}           & \multicolumn{1}{c|}{0.909}          & \textbf{0.942} \\ \cline{2-10} 
					& PBI-L                   & \multicolumn{1}{c|}{0.596}          & \multicolumn{1}{c|}{0.288}          & \multicolumn{1}{c|}{0.869}          & 0.706          & \multicolumn{1}{c|}{0.596}          & \multicolumn{1}{c|}{0.288}          & \multicolumn{1}{c|}{0.869}          & 0.706          \\ \cline{2-10} 
					& PBI-NL                  & \multicolumn{1}{c|}{0.43}           & \multicolumn{1}{c|}{0.248}          & \multicolumn{1}{c|}{0.851}          & 0.674          & \multicolumn{1}{c|}{0.43}           & \multicolumn{1}{c|}{0.248}          & \multicolumn{1}{c|}{0.851}          & 0.674          \\ \cline{2-10} 
					& HPDI                    & \multicolumn{1}{c|}{\textbf{0.993}} & \multicolumn{1}{c|}{\textbf{0.963}} & \multicolumn{1}{c|}{\textbf{0.974}} & \textbf{0.955} & \multicolumn{1}{c|}{\textbf{0.996}} & \multicolumn{1}{c|}{\textbf{0.96}}  & \multicolumn{1}{c|}{\textbf{0.942}} & 0.939          \\ \hline
					\multirow{4}{*}{Scheme-III}                                                 & MACI                    & \multicolumn{1}{c|}{0.905}          & \multicolumn{1}{c|}{0.913}          & \multicolumn{1}{c|}{0.933}          & 0.947          & \multicolumn{1}{c|}{0.916}          & \multicolumn{1}{c|}{0.918}          & \multicolumn{1}{c|}{0.889}          & 0.942          \\ \cline{2-10} 
					& PBI-L                   & \multicolumn{1}{c|}{0.1}            & \multicolumn{1}{c|}{0.503}          & \multicolumn{1}{c|}{0.542}          & 0.804          & \multicolumn{1}{c|}{0.1}            & \multicolumn{1}{c|}{0.503}          & \multicolumn{1}{c|}{0.804}          & 0.804          \\ \cline{2-10} 
					& PBI-NL                  & \multicolumn{1}{c|}{0.076}          & \multicolumn{1}{c|}{0.142}          & \multicolumn{1}{c|}{0.486}          & 0.663          & \multicolumn{1}{c|}{0.076}          & \multicolumn{1}{c|}{0.142}          & \multicolumn{1}{c|}{0.486}          & 0.663          \\ \cline{2-10} 
					& HPDI                    & \multicolumn{1}{c|}{\textbf{0.992}} & \multicolumn{1}{c|}{\textbf{0.97}}  & \multicolumn{1}{c|}{\textbf{0.977}} & \textbf{0.962} & \multicolumn{1}{c|}{\textbf{0.995}} & \multicolumn{1}{c|}{\textbf{0.963}} & \multicolumn{1}{c|}{\textbf{0.961}} & \textbf{0.953} \\ \hline
					
					\multicolumn{10}{|c|}{$\tau= 0.525$} \\
					\hline
					\multirow{3}{*}{\begin{tabular}[c]{@{}c@{}}Sampling \\ Scheme\end{tabular}} & \multirow{3}{*}{Method}
					& \multicolumn{4}{c|}{CV$_p$}
					& \multicolumn{4}{c|}{CV$_k$} \\ \cline{3-10}
					& & \multicolumn{2}{c|}{$n=50$} & \multicolumn{2}{c|}{$n=200$} & \multicolumn{2}{c|}{$n=50$} & \multicolumn{2}{c|}{$n=200$} \\ \cline{3-10}
					& & \multicolumn{1}{c|}{$m=4$} & \multicolumn{1}{c|}{$m=8$} & \multicolumn{1}{c|}{$m=4$} & $m=8$
					& \multicolumn{1}{c|}{$m=4$} & \multicolumn{1}{c|}{$m=8$} & \multicolumn{1}{c|}{$m=4$} & $m=8$ \\ \hline
					
					\multirow{4}{*}{Scheme-I}                                                   & MACI                    & \multicolumn{1}{c|}{0.942}          & \multicolumn{1}{c|}{0.949}          & \multicolumn{1}{c|}{0.954}          & 0.946          & \multicolumn{1}{c|}{0.938}          & \multicolumn{1}{c|}{0.939}          & \multicolumn{1}{c|}{0.942}          & 0.949          \\ \cline{2-10} 
					& PBI-L                   & \multicolumn{1}{c|}{0.912}          & \multicolumn{1}{c|}{0.868}          & \multicolumn{1}{c|}{0.938}          & 0.909          & \multicolumn{1}{c|}{0.912}          & \multicolumn{1}{c|}{0.868}          & \multicolumn{1}{c|}{0.938}          & 0.909          \\ \cline{2-10} 
					& PBI-NL                  & \multicolumn{1}{c|}{0.882}          & \multicolumn{1}{c|}{0.859}          & \multicolumn{1}{c|}{0.939}          & 0.914          & \multicolumn{1}{c|}{0.882}          & \multicolumn{1}{c|}{0.859}          & \multicolumn{1}{c|}{0.939}          & 0.914          \\ \cline{2-10} 
					& HPDI                    & \multicolumn{1}{c|}{\textbf{0.956}} & \multicolumn{1}{c|}{\textbf{0.97}}  & \multicolumn{1}{c|}{\textbf{0.958}} & \textbf{0.95}  & \multicolumn{1}{c|}{\textbf{0.943}} & \multicolumn{1}{c|}{\textbf{0.945}} & \multicolumn{1}{c|}{\textbf{0.945}} & \textbf{0.946} \\ \hline
					\multirow{4}{*}{Scheme-II}                                                  & MACI                    & \multicolumn{1}{c|}{0.937}          & \multicolumn{1}{c|}{0.98}           & \multicolumn{1}{c|}{0.944}          & 0.962          & \multicolumn{1}{c|}{0.929}          & \multicolumn{1}{c|}{0.96}           & \multicolumn{1}{c|}{0.947}          & 0.954          \\ \cline{2-10} 
					& PBI& \multicolumn{1}{c|}{0.919}          & \multicolumn{1}{c|}{0.931}          & \multicolumn{1}{c|}{0.937}          & 0.948          & \multicolumn{1}{c|}{0.919}          & \multicolumn{1}{c|}{0.931}          & \multicolumn{1}{c|}{0.937}          & 0.948          \\ \cline{2-10} 
					& PBI-NL                  & \multicolumn{1}{c|}{0.891}          & \multicolumn{1}{c|}{0.902}          & \multicolumn{1}{c|}{0.923}          & 0.955          & \multicolumn{1}{c|}{0.891}          & \multicolumn{1}{c|}{0.902}          & \multicolumn{1}{c|}{0.923}          & \textbf{0.955} \\  \cline{2-10}
					& HPDI                    & \multicolumn{1}{c|}{\textbf{0.965}} & \multicolumn{1}{c|}{\textbf{0.991}} & \multicolumn{1}{c|}{\textbf{0.954}} & \textbf{0.968} & \multicolumn{1}{c|}{\textbf{0.952}} & \multicolumn{1}{c|}{\textbf{0.967}} & \multicolumn{1}{c|}{\textbf{0.949}} & 0.956          \\\hline
					\multirow{4}{*}{Scheme-III}                                                 & MACI                    & \multicolumn{1}{c|}{0.945}          & \multicolumn{1}{c|}{0.956}          & \multicolumn{1}{c|}{0.951}          & 0.946          & \multicolumn{1}{c|}{0.927}          & \multicolumn{1}{c|}{0.945}          & \multicolumn{1}{c|}{0.933}          & 0.956          \\ \cline{2-10} 
					& PBI-L                   & \multicolumn{1}{c|}{0.885}          & \multicolumn{1}{c|}{0.889}          & \multicolumn{1}{c|}{0.906}          & 0.92           & \multicolumn{1}{c|}{0.885}          & \multicolumn{1}{c|}{0.889}          & \multicolumn{1}{c|}{0.906}          & 0.92           \\ \cline{2-10} 
					& PBI-NL                  & \multicolumn{1}{c|}{0.954}          & \multicolumn{1}{c|}{0.909}          & \multicolumn{1}{c|}{0.961}          & \textbf{0.958} & \multicolumn{1}{c|}{0.954}          & \multicolumn{1}{c|}{0.909}          & \multicolumn{1}{c|}{\textbf{0.961}} & \textbf{0.958} \\ \cline{2-10} 
					& HPDI                    & \multicolumn{1}{c|}{\textbf{0.988}} & \multicolumn{1}{c|}{\textbf{0.973}} & \multicolumn{1}{c|}{\textbf{0.962}} & 0.957          & \multicolumn{1}{c|}{\textbf{0.961}} & \multicolumn{1}{c|}{\textbf{0.955}} & \multicolumn{1}{c|}{0.948}          & \textbf{0.958} \\        
				\end{longtable}
			\end{center}
			\vspace{-1cm}
			 the estimators of the CV$_k$ perform better when the setting shape is smaller and the setting rate parameter is high.
			\item Let $\kappa$ be fixed. Regarding setting values of $\tau$, the estimators of $\kappa$ when $\tau=0.525$ generally perform better than when $\tau=0.052$. The opposite is observed for the estimators of $\tau$. 
			The estimators of the CV$_p$, perform better for $\tau=0.525$ when $\kappa=0.75$, and better for $\tau=0.052$ when $\kappa=1.25$. Furthermore, when $\kappa=0.75$, the estimators of the CV$_k$ have lower MSEs when $\tau=0.525$.
			\item In terms of the estimators, the Bayesian estimator performs the best in almost all the cases. The proposed least squares estimators also perform relatively well for the CV$_p$ in almost all cases.
			\end{enumerate}
			\indent Now, the interval estimators are compared based on their CPs and AWs. The CPs and AWs of various interval estimates of the CV$_p$ at 95\% confidence level are given in Table \ref{tab:my-table-3}-\ref{tab:my-table-7}. Based entirely on the methods, the comparisons are made for each parameter as follows:
			\begin{center}
	\footnotesize
	\begin{longtable}[H]{|c|c|cccc|cccc|}
		\caption{Coverage Probabilities of the interval estimates for 95\% confidence of CV$_p$ and CV$_k$ under various settings with $\kappa = 1.250$} \label{tab:my-table-5} \\ \hline
		\multicolumn{10}{|c|}{$\tau= 0.052$} \\
		\hline
		\multirow{3}{*}{\begin{tabular}[c]{@{}c@{}}Sampling \\ Scheme\end{tabular}}
		& \multirow{3}{*}{Method}
		& \multicolumn{4}{c|}{CV$_p$}
		& \multicolumn{4}{c|}{CV$_k$}
		\\ \cline{3-10}
		& & \multicolumn{2}{c|}{$n=50$} & \multicolumn{2}{c|}{$n=200$}
		& \multicolumn{2}{c|}{$n=50$} & \multicolumn{2}{c|}{$n=200$}
		\\ \cline{3-10}
		& & \multicolumn{1}{c|}{$m=4$} & \multicolumn{1}{c|}{$m=8$} & \multicolumn{1}{c|}{$m=4$} & $m=8$
		& \multicolumn{1}{c|}{$m=4$} & \multicolumn{1}{c|}{$m=8$} & \multicolumn{1}{c|}{$m=4$} & $m=8$
		\\ \hline
		\endfirsthead
		
		\hline
		\multicolumn{10}{|l|}{\textit{(continued from previous page)}} \\
		\hline
		\endhead
		
		\hline
		\multicolumn{10}{|r|}{\textit{Continued on next page}} \\
		\hline
		\endfoot
		
		\hline
		\endlastfoot
		\multirow{4}{*}{Scheme-I}                                                   & MACI                    & \multicolumn{1}{c|}{0.955}          & \multicolumn{1}{c|}{0.94}           & \multicolumn{1}{c|}{0.937}          & \textbf{0.955} & \multicolumn{1}{c|}{0.932}          & \multicolumn{1}{c|}{0.942}          & \multicolumn{1}{c|}{0.931}          & 0.95           \\ \cline{2-10} 
		& PBI-L                   & \multicolumn{1}{c|}{0.973}          & \multicolumn{1}{c|}{\textbf{0.968}} & \multicolumn{1}{c|}{\textbf{0.956}} & 0.954          & \multicolumn{1}{c|}{0.973}          & \multicolumn{1}{c|}{\textbf{0.968}} & \multicolumn{1}{c|}{\textbf{0.956}} & \textbf{0.954} \\ \cline{2-10} 
		& PBI-NL                  & \multicolumn{1}{c|}{0.964}          & \multicolumn{1}{c|}{0.888}          & \multicolumn{1}{c|}{0.936}          & 0.928          & \multicolumn{1}{c|}{0.964}          & \multicolumn{1}{c|}{0.888}          & \multicolumn{1}{c|}{0.936}          & 0.928          \\ \cline{2-10} 
		& HPDI                    & \multicolumn{1}{c|}{\textbf{0.987}} & \multicolumn{1}{c|}{0.967}          & \multicolumn{1}{c|}{0.943}          & 0.952          & \multicolumn{1}{c|}{\textbf{0.976}} & \multicolumn{1}{c|}{0.956}          & \multicolumn{1}{c|}{0.938}          & 0.947          \\ \hline
		\multirow{4}{*}{Scheme-II}                                                  & MACI                    & \multicolumn{1}{c|}{0.966}          & \multicolumn{1}{c|}{0.96}           & \multicolumn{1}{c|}{0.963}          & 0.946          & \multicolumn{1}{c|}{0.935}          & \multicolumn{1}{c|}{0.955}          & \multicolumn{1}{c|}{0.948}          & \textbf{0.947} \\ \cline{2-10} 
		& PBI-L                   & \multicolumn{1}{c|}{0.945}          & \multicolumn{1}{c|}{0.952}          & \multicolumn{1}{c|}{0.954}          & 0.929          & \multicolumn{1}{c|}{0.945}          & \multicolumn{1}{c|}{0.952}          & \multicolumn{1}{c|}{0.954}          & 0.929          \\ \cline{2-10} 
		& PBI-NL                  & \multicolumn{1}{c|}{0.97}           & \multicolumn{1}{c|}{0.918}          & \multicolumn{1}{c|}{0.956}          & 0.93           & \multicolumn{1}{c|}{0.97}           & \multicolumn{1}{c|}{0.918}          & \multicolumn{1}{c|}{\textbf{0.956}} & 0.93           \\ \cline{2-10} 
		& HPDI                    & \multicolumn{1}{c|}{\textbf{0.992}} & \multicolumn{1}{c|}{\textbf{0.977}} & \multicolumn{1}{c|}{\textbf{0.966}} & \textbf{0.947} & \multicolumn{1}{c|}{\textbf{0.975}} & \multicolumn{1}{c|}{\textbf{0.971}} & \multicolumn{1}{c|}{0.953}          & \textbf{0.947} \\ \hline
		\multirow{4}{*}{Scheme-III}                                                 & MACI                    & \multicolumn{1}{c|}{0.952}          & \multicolumn{1}{c|}{0.948}          & \multicolumn{1}{c|}{0.949}          & 0.952          & \multicolumn{1}{c|}{0.923}          & \multicolumn{1}{c|}{0.939}          & \multicolumn{1}{c|}{0.928}          & 0.948          \\ \cline{2-10} 
		& PBI-L                   & \multicolumn{1}{c|}{0.884}          & \multicolumn{1}{c|}{0.956}          & \multicolumn{1}{c|}{0.898}          & \textbf{0.958} & \multicolumn{1}{c|}{0.884}          & \multicolumn{1}{c|}{0.956}          & \multicolumn{1}{c|}{0.898}          & \textbf{0.958} \\ \cline{2-10} 
		& PBI-NL                  & \multicolumn{1}{c|}{0.788}          & \multicolumn{1}{c|}{0.836}          & \multicolumn{1}{c|}{0.847}          & 0.912          & \multicolumn{1}{c|}{0.788}          & \multicolumn{1}{c|}{0.836}          & \multicolumn{1}{c|}{0.847}          & 0.912          \\ \cline{2-10} 
		& HPDI                    & \multicolumn{1}{c|}{\textbf{0.991}} & \multicolumn{1}{c|}{\textbf{0.97}}  & \multicolumn{1}{c|}{\textbf{0.975}} & 0.955          & \multicolumn{1}{c|}{\textbf{0.991}} & \multicolumn{1}{c|}{\textbf{0.96}}  & \multicolumn{1}{c|}{\textbf{0.951}} & 0.952          \\ \hline
		
		\multicolumn{10}{|c|}{$\tau= 0.525$} \\
		\hline
		\multirow{3}{*}{\begin{tabular}[c]{@{}c@{}}Sampling \\ Scheme\end{tabular}} & \multirow{3}{*}{Method}
		& \multicolumn{4}{c|}{CV$_p$}
		& \multicolumn{4}{c|}{CV$_k$} \\ \cline{3-10}
		& & \multicolumn{2}{c|}{$n=50$} & \multicolumn{2}{c|}{$n=200$} & \multicolumn{2}{c|}{$n=50$} & \multicolumn{2}{c|}{$n=200$} \\ \cline{3-10}
		& & \multicolumn{1}{c|}{$m=4$} & \multicolumn{1}{c|}{$m=8$} & \multicolumn{1}{c|}{$m=4$} & $m=8$
		& \multicolumn{1}{c|}{$m=4$} & \multicolumn{1}{c|}{$m=8$} & \multicolumn{1}{c|}{$m=4$} & $m=8$ \\ \hline
		
		\multirow{4}{*}{Scheme-I}                                                   & MACI                    & \multicolumn{1}{c|}{0.954}          & \multicolumn{1}{c|}{0.836}          & \multicolumn{1}{c|}{0.942}          & 0.896          & \multicolumn{1}{c|}{0.943}          & \multicolumn{1}{c|}{0.741}          & \multicolumn{1}{c|}{0.941}          & 0.861          \\ \cline{2-10} 
		& PBI-L                   & \multicolumn{1}{c|}{0.968}          & \multicolumn{1}{c|}{0.259}          & \multicolumn{1}{c|}{0.949}          & 0.023          & \multicolumn{1}{c|}{\textbf{0.968}} & \multicolumn{1}{c|}{0.259}          & \multicolumn{1}{c|}{\textbf{0.949}} & 0.023          \\ \cline{2-10} 
		& PBI-NL                  & \multicolumn{1}{c|}{0.962}          & \multicolumn{1}{c|}{\textbf{0.95}}  & \multicolumn{1}{c|}{0.943}          & \textbf{0.919} & \multicolumn{1}{c|}{0.962}          & \multicolumn{1}{c|}{\textbf{0.95}}  & \multicolumn{1}{c|}{0.943}          & \textbf{0.919} \\ \cline{2-10} 
		& HPDI                    & \multicolumn{1}{c|}{\textbf{0.969}} & \multicolumn{1}{c|}{0.846}          & \multicolumn{1}{c|}{\textbf{0.952}} & 0.894          & \multicolumn{1}{c|}{0.96}           & \multicolumn{1}{c|}{0.753}          & \multicolumn{1}{c|}{0.944}          & 0.87           \\ \hline
		\multirow{4}{*}{Scheme-II}                                                  & MACI                    & \multicolumn{1}{c|}{0.983}          & \multicolumn{1}{c|}{0.735}          & \multicolumn{1}{c|}{0.954}          & 0.824          & \multicolumn{1}{c|}{0.969}          & \multicolumn{1}{c|}{0.492}          & \multicolumn{1}{c|}{0.956}          & 0.772          \\ \cline{2-10} 
		& PBI-L                   & \multicolumn{1}{c|}{0.948}          & \multicolumn{1}{c|}{0.125}          & \multicolumn{1}{c|}{\textbf{0.965}} & 0.001          & \multicolumn{1}{c|}{0.948}          & \multicolumn{1}{c|}{0.125}          & \multicolumn{1}{c|}{\textbf{0.965}} & 0.001          \\ \cline{2-10} 
		& PBI-NL                  & \multicolumn{1}{c|}{0.968}          & \multicolumn{1}{c|}{\textbf{0.814}} & \multicolumn{1}{c|}{0.964}          & \textbf{0.836} & \multicolumn{1}{c|}{0.968}          & \multicolumn{1}{c|}{\textbf{0.814}} & \multicolumn{1}{c|}{0.964}          & \textbf{0.836} \\ \cline{2-10} 
		& HPDI                    & \multicolumn{1}{c|}{\textbf{0.993}} & \multicolumn{1}{c|}{0.713}          & \multicolumn{1}{c|}{0.964}          & 0.823          & \multicolumn{1}{c|}{\textbf{0.979}} & \multicolumn{1}{c|}{0.493}          & \multicolumn{1}{c|}{0.964}          & 0.765          \\ \hline
		\multirow{4}{*}{Scheme-III}                                                 & MACI                    & \multicolumn{1}{c|}{\textbf{1}}     & \multicolumn{1}{c|}{0.802}          & \multicolumn{1}{c|}{0.967}          & 0.896          & \multicolumn{1}{c|}{0.926}          & \multicolumn{1}{c|}{0.651}          & \multicolumn{1}{c|}{0.953}          & 0.862          \\ \cline{2-10} 
		& PBI-L                   & \multicolumn{1}{c|}{0.863}          & \multicolumn{1}{c|}{0.193}          & \multicolumn{1}{c|}{0.928}          & 0              & \multicolumn{1}{c|}{0.863}          & \multicolumn{1}{c|}{0.193}          & \multicolumn{1}{c|}{0.928}          & 0              \\ \cline{2-10} 
		& PBI-NL                  & \multicolumn{1}{c|}{0.93}           & \multicolumn{1}{c|}{\textbf{0.924}} & \multicolumn{1}{c|}{0.933}          & \textbf{0.92}  & \multicolumn{1}{c|}{0.93}           & \multicolumn{1}{c|}{\textbf{0.924}} & \multicolumn{1}{c|}{0.933}          & \textbf{0.92}  \\ \cline{2-10} 
		& HPDI                    & \multicolumn{1}{c|}{0.997}          & \multicolumn{1}{c|}{0.811}          & \multicolumn{1}{c|}{\textbf{0.974}} & 0.896          & \multicolumn{1}{c|}{\textbf{0.953}} & \multicolumn{1}{c|}{0.673}          & \multicolumn{1}{c|}{\textbf{0.961}} & 0.867  
	\end{longtable}
\end{center}
\vspace{-1.5cm}
		\begin{enumerate}
			\item For $\kappa$: When $\kappa=0.75$, the HPDIs have the highest CPs in almost all cases. When $\kappa=1.25$ and $\tau=0.052$, the MACIs, PBI-L, and HPDIs have equivalently strong 
				performances as their CPs are close to or higher than the nominal values. When $\kappa=1.25$ and $\tau=0.525$, the PBI-NL performs best in almost all cases. Furthermore, the PBI-L and HPDIs have the smallest AWs.
			\item For $\tau$: The HPDI and MACI have the highest CPs in almost all cases. The PBIs do not have high CPs when the setting rate parameter is high. Despite this, the PBIs and the HPDIs have the smallest AWs.
			\item For CV$_p$ and CV$_k$, the HPDIs have the highest CPs in almost all the cases. Furthermore, the PBIs have the smallest AWs in almost all cases, with just a few minor exceptions.
			\end{enumerate}
				\begin{center} \footnotesize
				\begin{longtable}[H]{|c|c|cccc|cccc|}
					\caption{Average Widths of the interval estimates for 95\% confidence of CV$_p$ and CV$_k$ under various settings with $\kappa = 0.750$} \label{tab:my-table-6} \\
					\hline
					\multicolumn{10}{|c|}{$\tau= 0.052$} \\
					\hline
					\multirow{3}{*}{\begin{tabular}[c]{@{}c@{}}Sampling \\ Scheme\end{tabular}}
					& \multirow{3}{*}{Method}
					& \multicolumn{4}{c|}{CV$_p$}
					& \multicolumn{4}{c|}{CV$_k$}
					\\ \cline{3-10}
					& & \multicolumn{2}{c|}{$n=50$} & \multicolumn{2}{c|}{$n=200$}
					& \multicolumn{2}{c|}{$n=50$} & \multicolumn{2}{c|}{$n=200$}
					\\ \cline{3-10}
					& & \multicolumn{1}{c|}{$m=4$} & \multicolumn{1}{c|}{$m=8$} & \multicolumn{1}{c|}{$m=4$} & $m=8$
					& \multicolumn{1}{c|}{$m=4$} & \multicolumn{1}{c|}{$m=8$} & \multicolumn{1}{c|}{$m=4$} & $m=8$
					\\ \hline
					\endfirsthead
					
					\hline
					\multicolumn{10}{|l|}{\textit{(continued from previous page)}} \\
					\hline
					\endhead
					
					\hline
					\multicolumn{10}{|r|}{\textit{Continued on next page}} \\
					\hline
					\endfoot
					
					\hline
					\endlastfoot
					
					\multirow{4}{*}{Scheme-I}                                                   & MACI                    & \multicolumn{1}{c|}{15.9927}         & \multicolumn{1}{c|}{3.6511}          & \multicolumn{1}{c|}{1.7735}          & 1.1062          & \multicolumn{1}{c|}{0.5601}          & \multicolumn{1}{c|}{0.4045}          & \multicolumn{1}{c|}{0.2893}          & 0.2052          \\ \cline{2-10} 
					& PBI-L                   & \multicolumn{1}{c|}{1.7370}          & \multicolumn{1}{c|}{0.8893}          & \multicolumn{1}{c|}{1.3942}          & 0.8269          & \multicolumn{1}{c|}{\textbf{0.3767}} & \multicolumn{1}{c|}{\textbf{0.2713}} & \multicolumn{1}{c|}{0.2654}          & 0.1941          \\ \cline{2-10} 
					& PBI-NL                  & \multicolumn{1}{c|}{\textbf{1.3232}} & \multicolumn{1}{c|}{\textbf{0.8675}} & \multicolumn{1}{c|}{\textbf{1.2550}} & \textbf{0.7771} & \multicolumn{1}{c|}{0.3883}          & \multicolumn{1}{c|}{0.3217}          & \multicolumn{1}{c|}{0.2632}          & 0.1920          \\ \cline{2-10} 
					& HPDI                    & \multicolumn{1}{c|}{4.4372}          & \multicolumn{1}{c|}{2.1734}          & \multicolumn{1}{c|}{1.8771}          & 1.0682          & \multicolumn{1}{c|}{0.3991}          & \multicolumn{1}{c|}{0.3041}          & \multicolumn{1}{c|}{\textbf{0.2603}} & \textbf{0.1868} \\ \hline
					\multirow{4}{*}{Scheme-II}                                                  & MACI                    & \multicolumn{1}{c|}{20.2693}         & \multicolumn{1}{c|}{5.7658}          & \multicolumn{1}{c|}{4.0775}          & 1.2924          & \multicolumn{1}{c|}{0.6275}          & \multicolumn{1}{c|}{0.4577}          & \multicolumn{1}{c|}{0.3325}          & 0.2360          \\ \cline{2-10} 
					& PBI-L                   & \multicolumn{1}{c|}{0.9784}          & \multicolumn{1}{c|}{0.6450}          & \multicolumn{1}{c|}{1.2514}          & 0.6929          & \multicolumn{1}{c|}{0.3558}          & \multicolumn{1}{c|}{\textbf{0.2543}} & \multicolumn{1}{c|}{0.2828}          & 0.1993          \\ \cline{2-10} 
					& PBI-NL                  & \multicolumn{1}{c|}{\textbf{0.8071}} & \multicolumn{1}{c|}{\textbf{0.6283}} & \multicolumn{1}{c|}{\textbf{1.1910}} & \textbf{0.6709} & \multicolumn{1}{c|}{\textbf{0.3291}} & \multicolumn{1}{c|}{0.2616}          & \multicolumn{1}{c|}{0.2747}          & \textbf{0.1972} \\ \cline{2-10} 
					& HPDI                    & \multicolumn{1}{c|}{3.7371}          & \multicolumn{1}{c|}{2.3729}          & \multicolumn{1}{c|}{2.3996}          & 1.2532          & \multicolumn{1}{c|}{0.3746}          & \multicolumn{1}{c|}{0.3291}          & \multicolumn{1}{c|}{\textbf{0.2710}} & 0.2100          \\ \hline
					\multirow{4}{*}{Scheme-III}                                                 & MACI                    & \multicolumn{1}{c|}{14.5343}         & \multicolumn{1}{c|}{5.4020}          & \multicolumn{1}{c|}{9.8417}          & 1.1990          & \multicolumn{1}{c|}{0.7582}          & \multicolumn{1}{c|}{0.4411}          & \multicolumn{1}{c|}{0.4088}          & 0.2282          \\ \cline{2-10} 
					& PBI-L                   & \multicolumn{1}{c|}{\textbf{0.6382}} & \multicolumn{1}{c|}{0.7338}          & \multicolumn{1}{c|}{\textbf{0.8218}} & 0.7366          & \multicolumn{1}{c|}{\textbf{0.3184}} & \multicolumn{1}{c|}{0.2634}          & \multicolumn{1}{c|}{\textbf{0.2843}} & 0.1988          \\ \cline{2-10} 
					& PBI-NL                  & \multicolumn{1}{c|}{0.6619}          & \multicolumn{1}{c|}{\textbf{0.6032}} & \multicolumn{1}{c|}{0.8239}          & \textbf{0.6580} & \multicolumn{1}{c|}{0.3556}          & \multicolumn{1}{c|}{\textbf{0.2629}} & \multicolumn{1}{c|}{0.3052}          & \textbf{0.1976} \\ \cline{2-10} 
					& HPDI                    & \multicolumn{1}{c|}{3.9573}          & \multicolumn{1}{c|}{2.3002}          & \multicolumn{1}{c|}{3.0369}          & 1.1617          & \multicolumn{1}{c|}{0.4042}          & \multicolumn{1}{c|}{0.3224}          & \multicolumn{1}{c|}{0.3077}          & 0.2045          \\ \hline
					
					\multicolumn{10}{|c|}{$\tau= 0.525$} \\
					\hline
					\multirow{3}{*}{\begin{tabular}[c]{@{}c@{}}Sampling \\ Scheme\end{tabular}} & \multirow{3}{*}{Method}
					& \multicolumn{4}{c|}{CV$_p$}
					& \multicolumn{4}{c|}{CV$_k$} \\ \cline{3-10}
					& & \multicolumn{2}{c|}{$n=50$} & \multicolumn{2}{c|}{$n=200$} & \multicolumn{2}{c|}{$n=50$} & \multicolumn{2}{c|}{$n=200$} \\ \cline{3-10}
					& & \multicolumn{1}{c|}{$m=4$} & \multicolumn{1}{c|}{$m=8$} & \multicolumn{1}{c|}{$m=4$} & $m=8$
					& \multicolumn{1}{c|}{$m=4$} & \multicolumn{1}{c|}{$m=8$} & \multicolumn{1}{c|}{$m=4$} & $m=8$ \\ \hline
					
					\multirow{4}{*}{Scheme-I}                                                   & MACI                    & \multicolumn{1}{c|}{1.4580} & \multicolumn{1}{c|}{1.0432} & \multicolumn{1}{c|}{0.6343} & 0.4854 & \multicolumn{1}{c|}{0.2484} & \multicolumn{1}{c|}{0.1974} & \multicolumn{1}{c|}{0.1262} & 0.0996 \\ \cline{2-10} 
					& PBI-L                   & \multicolumn{1}{c|}{1.0908} & \multicolumn{1}{c|}{\textbf{0.6989}} & \multicolumn{1}{c|}{0.6034} & \textbf{0.4446} & \multicolumn{1}{c|}{0.2281} & \multicolumn{1}{c|}{\textbf{0.1732}} & \multicolumn{1}{c|}{0.1241} & \textbf{0.0970} \\ \cline{2-10} 
					& PBI-NL                  & \multicolumn{1}{c|}{\textbf{1.0531}} & \multicolumn{1}{c|}{0.7897} & \multicolumn{1}{c|}{\textbf{0.6025}} & 0.5102 & \multicolumn{1}{c|}{0.2373} & \multicolumn{1}{c|}{0.2051} & \multicolumn{1}{c|}{0.1269} & 0.1142 \\ \cline{2-10} 
					& HPDI                    & \multicolumn{1}{c|}{1.4105} & \multicolumn{1}{c|}{1.0356} & \multicolumn{1}{c|}{0.6329} & 0.4852 & \multicolumn{1}{c|}{\textbf{0.2208}} & \multicolumn{1}{c|}{0.1822} & \multicolumn{1}{c|}{\textbf{0.1221}} & 0.0974 \\ \hline
					\multirow{4}{*}{Scheme-II}                                                  & MACI                    & \multicolumn{1}{c|}{2.1074} & \multicolumn{1}{c|}{1.3681} & \multicolumn{1}{c|}{0.7901} & 0.6011 & \multicolumn{1}{c|}{0.3019} & \multicolumn{1}{c|}{0.2362} & \multicolumn{1}{c|}{0.1561} & 0.1210 \\ \cline{2-10} 
					& PBI-L                   & \multicolumn{1}{c|}{1.2162} & \multicolumn{1}{c|}{\textbf{0.7753}} & \multicolumn{1}{c|}{0.7253} & \textbf{0.5376} & \multicolumn{1}{c|}{\textbf{0.2545}} & \multicolumn{1}{c|}{\textbf{0.1858}} & \multicolumn{1}{c|}{\textbf{0.1488}} & \textbf{0.1143} \\ \cline{2-10} 
					& PBI-NL                  & \multicolumn{1}{c|}{\textbf{1.1552}} & \multicolumn{1}{c|}{0.8418} & \multicolumn{1}{c|}{\textbf{0.7144}} & 0.6008 & \multicolumn{1}{c|}{0.2672} & \multicolumn{1}{c|}{0.2159} & \multicolumn{1}{c|}{0.1521} & 0.1312 \\ \cline{2-10} 
					& HPDI                    & \multicolumn{1}{c|}{1.8984} & \multicolumn{1}{c|}{1.3941} & \multicolumn{1}{c|}{0.7986} & 0.6068 & \multicolumn{1}{c|}{0.2567} & \multicolumn{1}{c|}{0.2120} & \multicolumn{1}{c|}{0.1491} & 0.1176 \\ \hline
					\multirow{4}{*}{Scheme-III}                                                 & MACI                    & \multicolumn{1}{c|}{3.5964} & \multicolumn{1}{c|}{1.1297} & \multicolumn{1}{c|}{1.0530} & 0.5371 & \multicolumn{1}{c|}{0.3647} & \multicolumn{1}{c|}{0.2099} & \multicolumn{1}{c|}{0.1927} & 0.1100 \\ \cline{2-10} 
					& PBI-L                   & \multicolumn{1}{c|}{\textbf{1.0439}} & \multicolumn{1}{c|}{\textbf{0.7168}} & \multicolumn{1}{c|}{\textbf{0.8409}} & \textbf{0.4815} & \multicolumn{1}{c|}{\textbf{0.2587}} & \multicolumn{1}{c|}{\textbf{0.1772}} & \multicolumn{1}{c|}{0.1818} & \textbf{ 0.1062} \\ \cline{2-10} 
					& PBI-NL                  & \multicolumn{1}{c|}{1.5298} & \multicolumn{1}{c|}{0.8783} & \multicolumn{1}{c|}{0.9471} & 0.5725 & \multicolumn{1}{c|}{0.2949} & \multicolumn{1}{c|}{0.2155} & \multicolumn{1}{c|}{0.1810} & 0.1214 \\ \cline{2-10} 
					& HPDI                    & \multicolumn{1}{c|}{2.5576} & \multicolumn{1}{c|}{1.1250} & \multicolumn{1}{c|}{1.0613} & 0.5373 & \multicolumn{1}{c|}{0.2891} & \multicolumn{1}{c|}{0.1922} & \multicolumn{1}{c|}{\textbf{0.1796}} & 0.1071 \\ 
				\end{longtable}
			\end{center}
				\vspace{-1.5cm}
		\indent When no prior information is available, the Jeffrey's joint prior density can be used, but this prior function is improper. This results in Bayesian estimates that have only slightly higher MSEs than those with gamma prior in almost all cases. Correspondingly, the resulting HPDIs have a weak performance and therefore cannot cover the true parameter every time as much as the ones with known prior information.
		\subsection{Illustrative example using Real Data}
			\indent In this subsection, we apply the different methodologies in this paper to estimation of the parameters $\kappa$, $\tau$, CV$_p$, and CV$_k$ using a real Weibull lifetime data published in \cite{PPC} and used by some authors like \cite{JL}, \cite{HKTN} and \cite{CC}.
					\begin{center} \footnotesize
			\begin{longtable}[T]{|c|c|cccc|cccc|}
				\caption{Average Widths of the interval estimates for 95\% confidence of CV$_p$ and CV$_k$ under various settings with $\kappa = 1.250$} \label{tab:my-table-7} \\
				\hline
				\multicolumn{10}{|c|}{$\tau= 0.052$} \\
				\hline
				\multirow{3}{*}{\begin{tabular}[c]{@{}c@{}}Sampling \\ Scheme\end{tabular}}
				& \multirow{3}{*}{Method}
				& \multicolumn{4}{c|}{CV$_p$}
				& \multicolumn{4}{c|}{CV$_k$}
				\\ \cline{3-10}
				& & \multicolumn{2}{c|}{$n=50$} & \multicolumn{2}{c|}{$n=200$}
				& \multicolumn{2}{c|}{$n=50$} & \multicolumn{2}{c|}{$n=200$}
				\\ \cline{3-10}
				& & \multicolumn{1}{c|}{$m=4$} & \multicolumn{1}{c|}{$m=8$} & \multicolumn{1}{c|}{$m=4$} & $m=8$
				& \multicolumn{1}{c|}{$m=4$} & \multicolumn{1}{c|}{$m=8$} & \multicolumn{1}{c|}{$m=4$} & $m=8$
				\\ \hline
				\endfirsthead
				
				\hline
				\multicolumn{10}{|l|}{\textit{(continued from previous page)}} \\
				\hline
				\endhead
				
				\hline
				\multicolumn{10}{|r|}{\textit{Continued on next page}} \\
				\hline
				\endfoot
				
				\hline
				\endlastfoot
				\multirow{4}{*}{Scheme-I}                                                   & MACI                    & \multicolumn{1}{c|}{1.1419}          & \multicolumn{1}{c|}{0.6088}          & \multicolumn{1}{c|}{0.4783}          & 0.2989          & \multicolumn{1}{c|}{0.4320}          & \multicolumn{1}{c|}{0.2747}          & \multicolumn{1}{c|}{0.2160}          & 0.1383          \\ \cline{2-10} 
				& PBI-L                   & \multicolumn{1}{c|}{0.8840}          & \multicolumn{1}{c|}{\textbf{0.5017}} & \multicolumn{1}{c|}{0.5008}          & 0.3189          & \multicolumn{1}{c|}{0.3348}          & \multicolumn{1}{c|}{\textbf{0.2290}} & \multicolumn{1}{c|}{0.2196}          & 0.1486          \\ \cline{2-10} 
				& PBI-NL                  & \multicolumn{1}{c|}{\textbf{0.7841}} & \multicolumn{1}{c|}{0.5175}          & \multicolumn{1}{c|}{0.4634}          & 0.2984          & \multicolumn{1}{c|}{0.3477}          & \multicolumn{1}{c|}{0.2711}          & \multicolumn{1}{c|}{0.2137}          & 0.1448          \\ \cline{2-10} 
				& HPDI                    & \multicolumn{1}{c|}{0.8446}          & \multicolumn{1}{c|}{0.5318}          & \multicolumn{1}{c|}{\textbf{0.4446}} & \textbf{0.2843} & \multicolumn{1}{c|}{\textbf{0.3269}} & \multicolumn{1}{c|}{0.2339}          & \multicolumn{1}{c|}{\textbf{0.1964}} & \textbf{0.1301} \\ \hline
				\multirow{4}{*}{Scheme-II}                                                  & MACI                    & \multicolumn{1}{c|}{2.8743}          & \multicolumn{1}{c|}{0.7309}          & \multicolumn{1}{c|}{0.5712}          & 0.3416          & \multicolumn{1}{c|}{0.4870}          & \multicolumn{1}{c|}{0.3209}          & \multicolumn{1}{c|}{0.2485}          & 0.1589          \\ \cline{2-10} 
				& PBI-L                   & \multicolumn{1}{c|}{0.7340}          & \multicolumn{1}{c|}{\textbf{0.4712}} & \multicolumn{1}{c|}{0.5208}          & 0.3195          & \multicolumn{1}{c|}{0.3242}          & \multicolumn{1}{c|}{\textbf{0.2271}} & \multicolumn{1}{c|}{0.2308}          & 0.1534          \\ \cline{2-10} 
				& PBI-NL                  & \multicolumn{1}{c|}{\textbf{0.6981}} & \multicolumn{1}{c|}{0.4718}          & \multicolumn{1}{c|}{\textbf{0.4881}} & \textbf{0.3084} & \multicolumn{1}{c|}{\textbf{0.3186}} & \multicolumn{1}{c|}{0.2435}          & \multicolumn{1}{c|}{\textbf{0.2183}} & \textbf{0.1488} \\ \cline{2-10} 
				& HPDI                    & \multicolumn{1}{c|}{1.0131}          & \multicolumn{1}{c|}{0.6236}          & \multicolumn{1}{c|}{0.5271}          & 0.3253          & \multicolumn{1}{c|}{0.3554}          & \multicolumn{1}{c|}{0.2646}          & \multicolumn{1}{c|}{0.2225}          & \textbf{0.1488} \\ \hline
				\multirow{4}{*}{Scheme-III}                                                 & MACI                    & \multicolumn{1}{c|}{4.5897}          & \multicolumn{1}{c|}{0.6717}          & \multicolumn{1}{c|}{0.8082}          & 0.3302          & \multicolumn{1}{c|}{0.6190}          & \multicolumn{1}{c|}{0.2985}          & \multicolumn{1}{c|}{0.3163}          & 0.1532          \\ \cline{2-10} 
				& PBI-L                   & \multicolumn{1}{c|}{\textbf{0.5649}} & \multicolumn{1}{c|}{0.4815}          & \multicolumn{1}{c|}{\textbf{0.4986}} & 0.3196          & \multicolumn{1}{c|}{\textbf{0.3034}} & \multicolumn{1}{c|}{0.2275}          & \multicolumn{1}{c|}{\textbf{0.2397}} & 0.1517          \\ \cline{2-10} 
				& PBI-NL                  & \multicolumn{1}{c|}{0.5900}          & \multicolumn{1}{c|}{\textbf{0.4265}} & \multicolumn{1}{c|}{0.5069}          & \textbf{0.2997} & \multicolumn{1}{c|}{0.3387}          & \multicolumn{1}{c|}{\textbf{0.2271}} & \multicolumn{1}{c|}{0.2567}          & 0.1468          \\ \cline{2-10} 
				& HPDI                    & \multicolumn{1}{c|}{1.1405}          & \multicolumn{1}{c|}{0.5787}          & \multicolumn{1}{c|}{0.6958}          & 0.3142          & \multicolumn{1}{c|}{0.4052}          & \multicolumn{1}{c|}{0.2498}          & \multicolumn{1}{c|}{0.2699}          & \textbf{0.1436} \\ \hline
				
				\multicolumn{10}{|c|}{$\tau= 0.525$} \\
				\hline
				\multirow{3}{*}{\begin{tabular}[c]{@{}c@{}}Sampling \\ Scheme\end{tabular}} & \multirow{3}{*}{Method}
				& \multicolumn{4}{c|}{CV$_p$}
				& \multicolumn{4}{c|}{CV$_k$} \\ \cline{3-10}
				& & \multicolumn{2}{c|}{$n=50$} & \multicolumn{2}{c|}{$n=200$} & \multicolumn{2}{c|}{$n=50$} & \multicolumn{2}{c|}{$n=200$} \\ \cline{3-10}
				& & \multicolumn{1}{c|}{$m=4$} & \multicolumn{1}{c|}{$m=8$} & \multicolumn{1}{c|}{$m=4$} & $m=8$
				& \multicolumn{1}{c|}{$m=4$} & \multicolumn{1}{c|}{$m=8$} & \multicolumn{1}{c|}{$m=4$} & $m=8$ \\ \hline
				\multirow{4}{*}{Scheme-I}                                                   & MACI                    & \multicolumn{1}{c|}{0.4990}          & \multicolumn{1}{c|}{0.5507}          & \multicolumn{1}{c|}{0.2343}          & 0.2270          & \multicolumn{1}{c|}{0.2201}          & \multicolumn{1}{c|}{0.1939}          & \multicolumn{1}{c|}{0.1093}          & 0.0985          \\ \cline{2-10} 
				& PBI-L                   & \multicolumn{1}{c|}{\textbf{0.4715}} & \multicolumn{1}{c|}{0.5292}          & \multicolumn{1}{c|}{0.2324}          & 0.2430          & \multicolumn{1}{c|}{\textbf{0.1975}} & \multicolumn{1}{c|}{\textbf{0.1594}} & \multicolumn{1}{c|}{\textbf{0.1068}} & \textbf{0.0849} \\ \cline{2-10} 
				& PBI-NL                  & \multicolumn{1}{c|}{0.5748}          & \multicolumn{1}{c|}{0.6468}          & \multicolumn{1}{c|}{0.3023}          & 0.3419          & \multicolumn{1}{c|}{0.2436}          & \multicolumn{1}{c|}{0.2450}          & \multicolumn{1}{c|}{0.1379}          & 0.1465          \\ \cline{2-10} 
				& HPDI                    & \multicolumn{1}{c|}{0.4736}          & \multicolumn{1}{c|}{\textbf{0.5198}} & \multicolumn{1}{c|}{\textbf{0.2315}} & \textbf{0.2241} & \multicolumn{1}{c|}{0.2032}          & \multicolumn{1}{c|}{0.1803}          & \multicolumn{1}{c|}{0.1069}          & 0.0965          \\ \hline
				\multirow{4}{*}{Scheme-II} & MACI                    & \multicolumn{1}{c|}{0.6506}          & \multicolumn{1}{c|}{0.8324}          & \multicolumn{1}{c|}{0.2847}          & 0.2945          & \multicolumn{1}{c|}{0.2694}          & \multicolumn{1}{c|}{0.2376}          & \multicolumn{1}{c|}{0.1328}          & 0.1198          \\ \cline{2-10}
				& PBI-L                   & \multicolumn{1}{c|}{\textbf{0.6035}} & \multicolumn{1}{c|}{\textbf{0.6781}} & \multicolumn{1}{c|}{\textbf{0.2820}} & 0.3272          & \multicolumn{1}{c|}{\textbf{0.2221}} & \multicolumn{1}{c|}{\textbf{0.1747}} & \multicolumn{1}{c|}{\textbf{0.1256}} & \textbf{0.0984} \\ \cline{2-10} 
				& PBI-NL                  & \multicolumn{1}{c|}{0.6883}          & \multicolumn{1}{c|}{0.8014}          & \multicolumn{1}{c|}{0.3680}          & 0.4547          & \multicolumn{1}{c|}{0.2678}          & \multicolumn{1}{c|}{0.2508}          & \multicolumn{1}{c|}{0.1639}          & 0.1733          \\ \cline{2-10} 
				& HPDI                    & \multicolumn{1}{c|}{0.6120}          & \multicolumn{1}{c|}{0.7585}          & \multicolumn{1}{c|}{0.2824}          & \textbf{0.2908} & \multicolumn{1}{c|}{0.2432}          & \multicolumn{1}{c|}{0.2139}          & \multicolumn{1}{c|}{0.1294}          & 0.1167          \\ \hline
				\multirow{4}{*}{Scheme-III}                                                 & MACI                    & \multicolumn{1}{c|}{1.0898}          & \multicolumn{1}{c|}{0.6031}          & \multicolumn{1}{c|}{0.3622}          & 0.2441          & \multicolumn{1}{c|}{0.3421}          & \multicolumn{1}{c|}{0.2025}          & \multicolumn{1}{c|}{0.1623}          & 0.1046          \\ \cline{2-10} 
				& PBI-L                   & \multicolumn{1}{c|}{\textbf{0.7952}} & \multicolumn{1}{c|}{\textbf{0.5705}} & \multicolumn{1}{c|}{0.3586}          & 0.2751          & \multicolumn{1}{c|}{\textbf{0.2431}} & \multicolumn{1}{c|}{\textbf{0.1624}} & \multicolumn{1}{c|}{\textbf{0.1439}} & \textbf{0.0895} \\ \cline{2-10} 
				& PBI-NL                  & \multicolumn{1}{c|}{1.0336}          & \multicolumn{1}{c|}{0.7145}          & \multicolumn{1}{c|}{0.4393}          & 0.3519          & \multicolumn{1}{c|}{0.3052}          & \multicolumn{1}{c|}{0.2532}          & \multicolumn{1}{c|}{0.1811}          & 0.1502          \\ \cline{2-10} 
				& HPDI                    & \multicolumn{1}{c|}{0.9088}          & \multicolumn{1}{c|}{0.5655}          & \multicolumn{1}{c|}{\textbf{0.3566}} & \textbf{0.2410} & \multicolumn{1}{c|}{0.2866}          & \multicolumn{1}{c|}{0.1873}          & \multicolumn{1}{c|}{0.1561}          & 0.1023    
			\end{longtable}
		\end{center}
		\vspace{-1.5cm}
	In this study, 112 patients were treated for plasma cell myeloma at the National Cancer Institute. The survival times in months were observed at ten intervals, namely, $\interval({0,5.5}]$,$\interval({5.5,10.5}],...,$$\interval({50.5,60.5}]$, $(60.5,\infty)$. The number of patients at risk and those who dropped out of the treatment had also been observed. That is, $\boldsymbol{X}=(18,16,18,10,...,4,1)$ and $\boldsymbol{W}=(1,1,3,0,...,3,2)$. We can summarise this problem in the Figure 1 below: 
	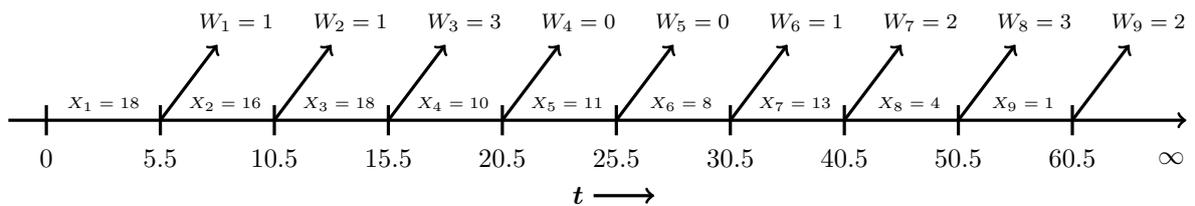
\begin{figure}[H]
		\footnotesize
		\begin{center}
			\begin{tikzpicture}
				[mydrawstyle/.style={draw=black, very thick}, x=1mm, y=1mm, z=1mm]
				\draw[mydrawstyle, ->](-15,40)--(140,40) node at (-8,40){};
				\draw[mydrawstyle](-10,42)--(-10,38) node[below=0.5]{$0$};
				\draw (-2.5,40) node[above] {\tiny{${X_1 = 18}$}};
				\draw[mydrawstyle](5,42)--(5,38) node[below=0.5]{$5.5$};
				\draw (13.5,40) node[above] {\tiny{$X_2=16$}};
				\draw[mydrawstyle](20,42)--(20,38) node[below=0.5]{$10.5$};
				\draw (28.5,40) node[above] {\tiny{$X_3=18$}};
				\draw[mydrawstyle](35,42)--(35,38) node[below=0.5]{$15.5$};
				\draw (43.5,40) node[above] {\tiny{$X_4=10$}};
				\draw[mydrawstyle](50,42)--(50,38) node[below=0.5]{$20.5$};
				\draw (58.5,40) node[above] {\tiny{$X_5=11$}};
				\draw[mydrawstyle](65,42)--(65,38) node[below=0.5]{$25.5$};
				\draw (73.5,40) node[above] {\tiny{$X_6=8$}};
				\draw[mydrawstyle](80,42)--(80,38) node[below=0.5]{$30.5$};
				\draw (88.5,40) node[above] {\tiny{$X_7=13$}};
				\draw[mydrawstyle](95,42)--(95,38) node[below=0.5]{$40.5$};
				\draw (103.5,40) node[above] {\tiny{$X_8=4$}};
				\draw[mydrawstyle](110,42)--(110,38) node[below=0.5]{$50.5$};
				\draw (118.5,40) node[above] {\tiny{$X_9=1$}};
				\draw[mydrawstyle](125,42)--(125,38) node[below=0.5]{$60.5$};
				\draw (138,37) node[below] {$\infty$};
				\draw[mydrawstyle, ->](62,30)--(70,30);
				\draw (60,30) node{$\boldsymbol{t}$};
				\draw[mydrawstyle, ->](5,40)--(12.5,50) node at (15,53){\scriptsize{$W_1=1$}};
				\draw[mydrawstyle, ->](20,40)--(27.5,50) node at (30,53){\scriptsize{$W_2=1$}};
				\draw[mydrawstyle, ->](35,40)--(42.5,50) node at (45,53){\scriptsize{$W_3=3$}};
				\draw[mydrawstyle, ->](50,40)--(57.5,50) node at (60,53){\scriptsize{$W_4=0$}};
				\draw[mydrawstyle, ->](65,40)--(72.5,50) node at (75,53){\scriptsize{$W_5=0$}};
				\draw[mydrawstyle, ->](80,40)--(87.5,50) node at (90,53){\scriptsize{$W_6=1$}};
				\draw[mydrawstyle, ->](95,40)--(102.5,50) node at (105,53){\scriptsize{$W_7=2$}};
				\draw[mydrawstyle, ->](110,40)--(117.5,50) node at (120,53){\scriptsize{$W_8=3$}};
				\draw[mydrawstyle, ->](125,40)--(132.5,50) node at (135,53){\scriptsize{$W_9=2$}};
			\end{tikzpicture}
			\label{figure}
			\captionof{figure}{Progressively Type-I Interval Censored Lifetime Data of Plasma Cell Myeloma Patients}
		\end{center}	
	\end{figure}
	\begin{table}[H]
		\centering
		\caption{Point Estimates of CV using Real dataset}
		\label{tab:my-table-8}
		\begin{tabular}{ccccc}
			\hline
			\multicolumn{1}{l}{} & MLE       & LLSE       & NLLSE      & Bayesian  \\ \hline
			$\kappa$               & 1.2297  & 1.2164   & 1.2352   & 1.2433\\ \hline
			$\tau$                 & 0.0211 & 0.0209 & 0.0199 & 0.0216\\ \hline
			CV$_p$                 & 0.8176 & 0.8261  & 0.8141  & 0.8163 \\ \hline
			CV$_k$                 & 0.6329 & 0.6369  & 0.6313  & 		0.6305 \\ \hline
		\end{tabular}
	\end{table}
		\begin{table}[h]
		\centering
		\caption{Interval estimates for $\kappa$ and $\tau$ at 95\% confidence level}
		\label{tab:my-table-9}
		\begin{tabular}{c|ccc|ccc}
			\hline
			\multirow{2}{*}{Methods} & \multicolumn{3}{c|}{$\kappa$}                                                            & \multicolumn{3}{c}{$\tau$}                                                              \\ \cline{2-7} 
			& \multicolumn{1}{c}{Lower Limit} & \multicolumn{1}{c}{Upper Limit} & Width           & \multicolumn{1}{c}{Lower Limit} & \multicolumn{1}{c}{Upper Limit} & Width           \\ \hline
			MACI                     & \multicolumn{1}{c}{1.0329}      & \multicolumn{1}{c}{1.4640}      & 0.4311          & \multicolumn{1}{c}{0.0100}      & \multicolumn{1}{c}{0.0443}      & 0.0343          \\ \hline
			PBI-L                     & \multicolumn{1}{c}{1.0367}      & \multicolumn{1}{c}{1.4360}      & \textbf{0.3993}          & \multicolumn{1}{c}{0.0091}      & \multicolumn{1}{c}{0.0383}      & \textbf{0.0292 }         \\ \hline
			PBI-NL                 & \multicolumn{1}{c}{1.0304}      & \multicolumn{1}{c}{1.5032}      & 0.4728          & \multicolumn{1}{c}{0.0080}      & \multicolumn{1}{c}{0.0388}      & 0.0308          \\ \hline
			HPDI              & \multicolumn{1}{c}{1.0336}      & \multicolumn{1}{c}{	1.4910}      & 	0.4574 & \multicolumn{1}{c}{0.0071}      & \multicolumn{1}{c}{		0.0391}      & 	0.0319 \\ \hline
		\end{tabular}
	\end{table}
	\begin{table}[h]
		\centering
		\caption{Interval estimates for CV$_p$ and CV$_k$ at 95\% confidence level}
		\label{tab:my-table-10}
		\begin{tabular}{c|ccc|ccc}
			\hline
			\multirow{2}{*}{Methods} & \multicolumn{3}{c|}{CV$_p$}                                                           & \multicolumn{3}{c}{CV$_k$}                                                           \\ \cline{2-7} 
			& \multicolumn{1}{c}{Lower Limit} & \multicolumn{1}{c}{Upper Limit} & Width           & \multicolumn{1}{c}{Lower Limit} & \multicolumn{1}{c}{Upper Limit} & Width           \\ \hline
			MACI                     & \multicolumn{1}{c}{0.6926}      & \multicolumn{1}{c}{0.9651}      & 0.2725          & \multicolumn{1}{c}{0.5731}      & \multicolumn{1}{c}{0.6991}      & 0.1261          \\ \hline
			PBI-L                      & \multicolumn{1}{c}{0.7062}      & \multicolumn{1}{c}{0.9615}      & \textbf{0.2552 }         & \multicolumn{1}{c}{0.5769}      & \multicolumn{1}{c}{0.6931}      & \textbf{0.1162  }        \\ \hline
			PBI-NL                 & \multicolumn{1}{c}{0.6767}      & \multicolumn{1}{c}{0.9702}      & 0.2935          & \multicolumn{1}{c}{0.5604}      & \multicolumn{1}{c}{0.6963}      & 0.1359          \\ \hline
			HPDI              & \multicolumn{1}{c}{0.6813}      & \multicolumn{1}{c}{	0.9648}      & 	0.2835 & \multicolumn{1}{c}{0.5630}      & \multicolumn{1}{c}{		0.6943}      & 	0.1313\\ \hline
		\end{tabular}
	\end{table}
		The various algorithms in this paper are applied. For the bootstrap estimation, we have considered a resampling of 2000 times. For the Bayesian estimation, we have created a large enough chain and also considered the process of thinning the chain by selecting one in every 100th generated sample to obtain a final MCMC sample of size $M=4500$. This process is practiced widely to avoid a high dependency on the previously generated samples. Furthermore, since no prior information is available, we have employed the non-informative Jeffrey prior density. The covariance matrix is set as small as diag$(5\times10^{-5})$ to achieve an acceptance rate of 34.5\%. \\
		\indent The results are displayed in Table \ref{tab:my-table-8}-\ref{tab:my-table-10}. It is observed that all the point estimates are almost consistent with their resulting values. Regarding the interval estimates, the PBIs obtained using the linear least squares estimate are the smallest in width in all cases. The intervals and their corresponding point estimates are displayed in Figure \ref{fig:2}. Here, the black dot represents the corresponding point estimate. For the least squares estimation methods, we have used this real dataset and plotted the surface plots using MATLAB R2023a in Figure \ref{fig:SSe}-\ref{fig:So}. The least squares fit using both methods are given in Figure \ref{LSF}-\ref{NLLSF}. In Figure \ref{fig:traceplots} and \ref{fig:Histogram}, we display some MCMC diagnostic results for the parameters. To obtain the histograms and HPDIs plot we have used the R code provided by \cite{JKK}.
				\begin{figure}[H]
			\centering
			\includegraphics[width=0.75\linewidth]{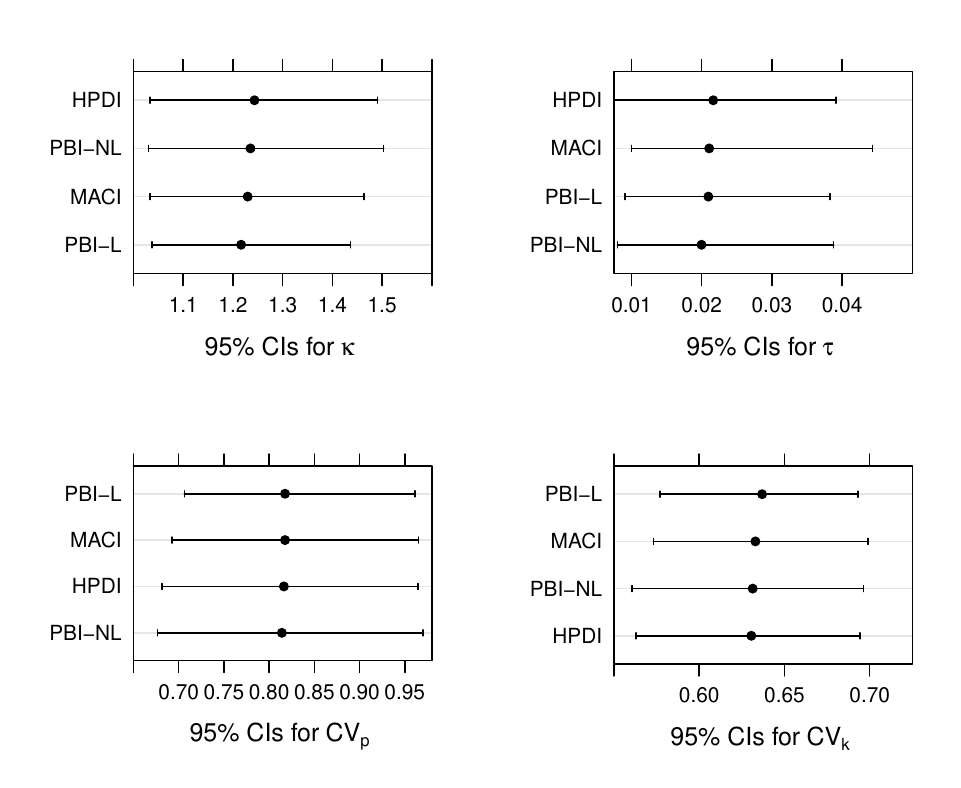}
			\captionof{figure}{The interval estimates of the various parameters using real data}
			\label{fig:2}
		\end{figure}
	\begin{figure}[H]
		\centering
		\begin{subfigure}[b]{0.45\textwidth}
			\centering
			\includegraphics[width=\textwidth]{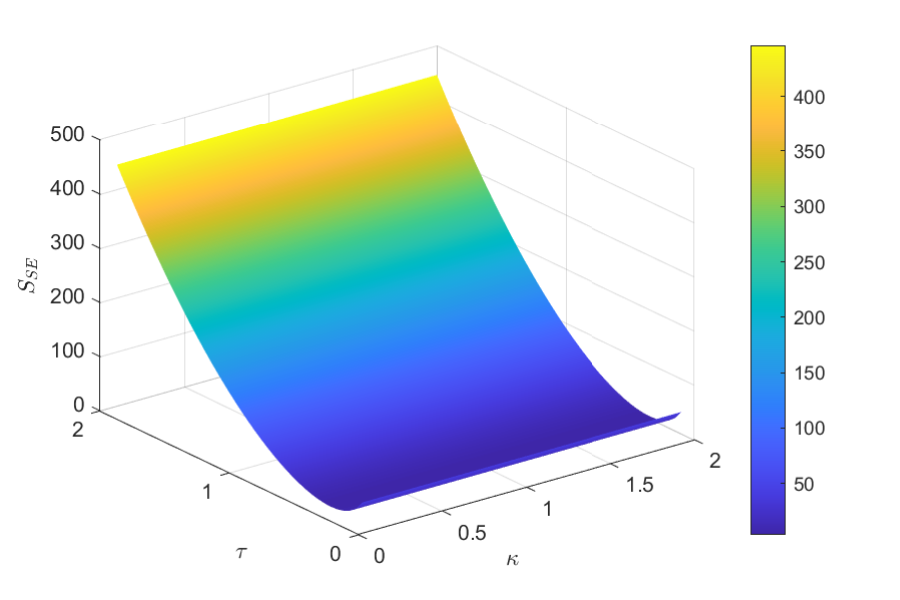}
			\caption{Surface Plot of S$_{SE}$}
			\label{fig:SSe}
		\end{subfigure}
		\hspace{0.75cm}
		\begin{subfigure}[b]{0.45\textwidth}
			\centering
			\includegraphics[width=\textwidth]{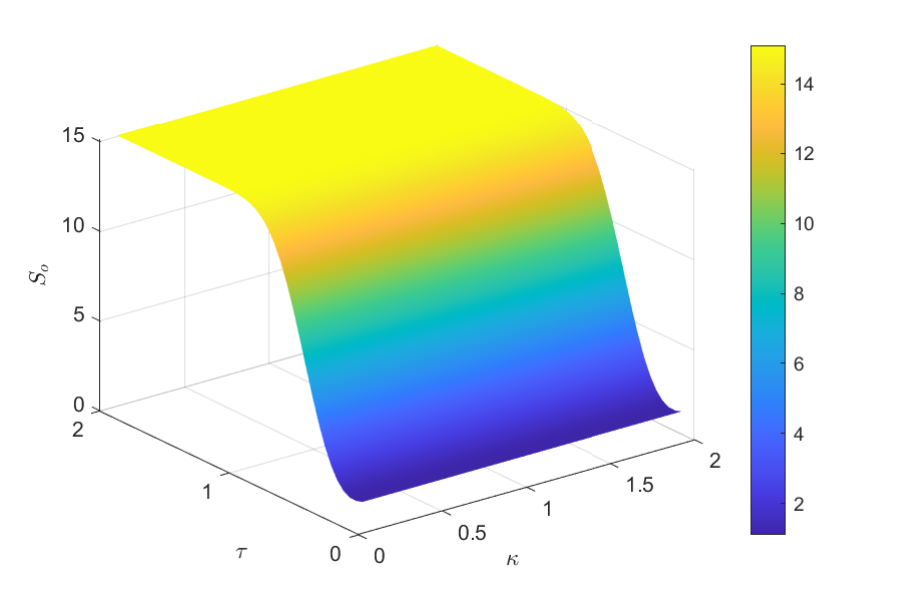}
			\caption{Surface Plot of S$_o$}
			\label{fig:So}
		\end{subfigure}
		\hfill
		\begin{subfigure}[b]{0.45\textwidth}
			\centering
			\includegraphics[width=\textwidth]{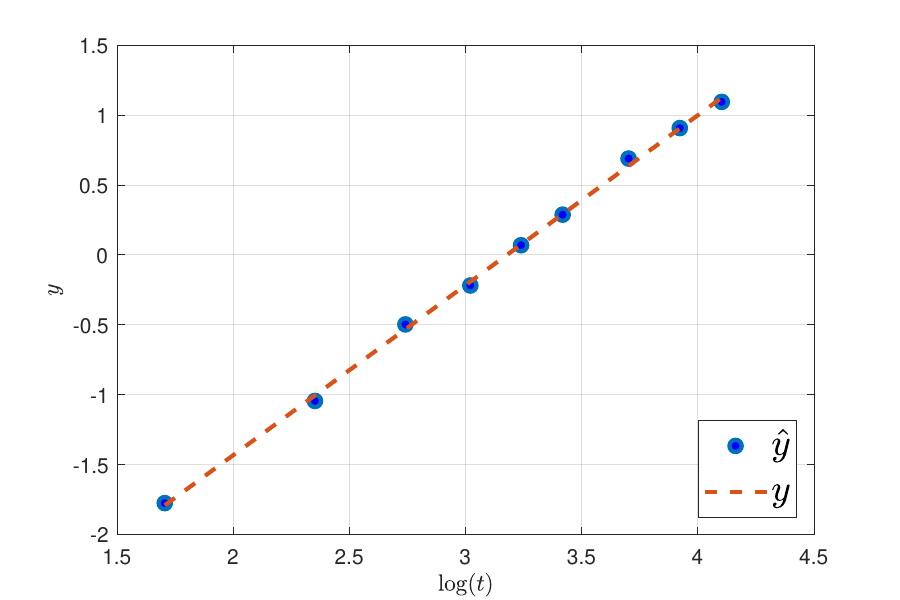}
			\caption{Fitted linear model vs $\log t$}
			\label{LSF}
		\end{subfigure}
		\hspace{0.75cm}
		\begin{subfigure}[b]{0.45\textwidth}
			\centering
			\includegraphics[width=\textwidth]{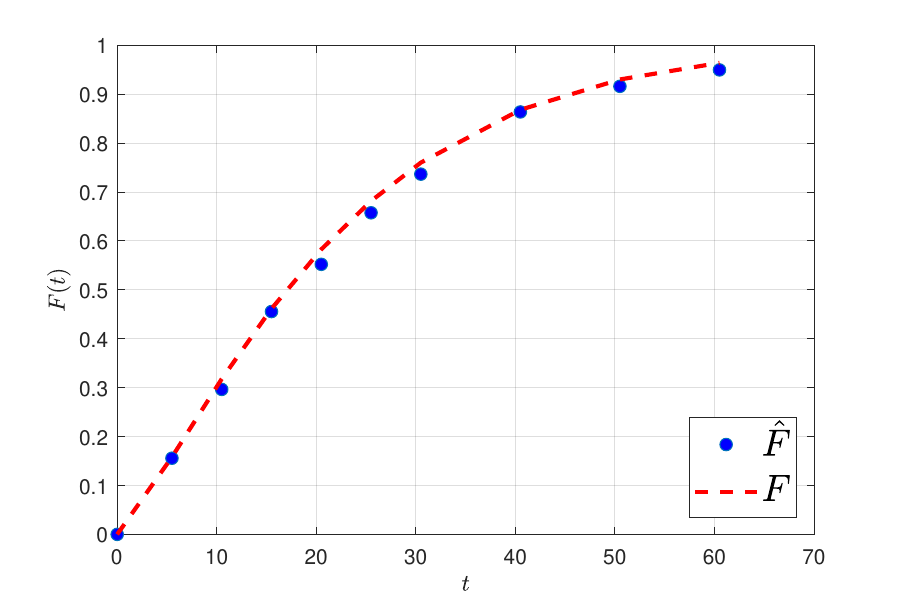}
			\caption{Fitted distribution function curve with non-parametric estimates vs $t$}
			\label{NLLSF}
		\end{subfigure}
		\caption{(a-b) Least Squares surface plots of the objective functions (c-d) curve fitting using real data}
		\label{Least Squares Fits}
	\end{figure}

		\begin{figure}
			\centering
			\begin{subfigure}[b]{0.35\textwidth}
				\centering
				\includegraphics[width=\textwidth]{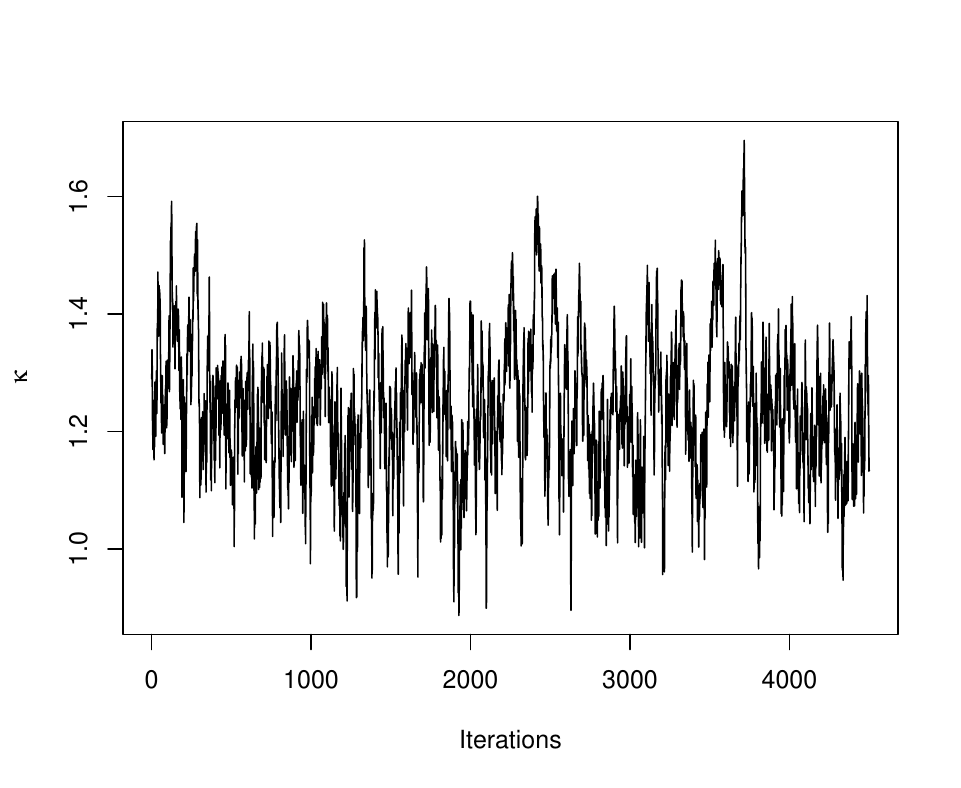}
				\caption{$\kappa$}
				\label{fig:kappa}
			\end{subfigure}
			\hspace{1cm}
			\begin{subfigure}[b]{0.35\textwidth}
				\centering
				\includegraphics[width=\textwidth]{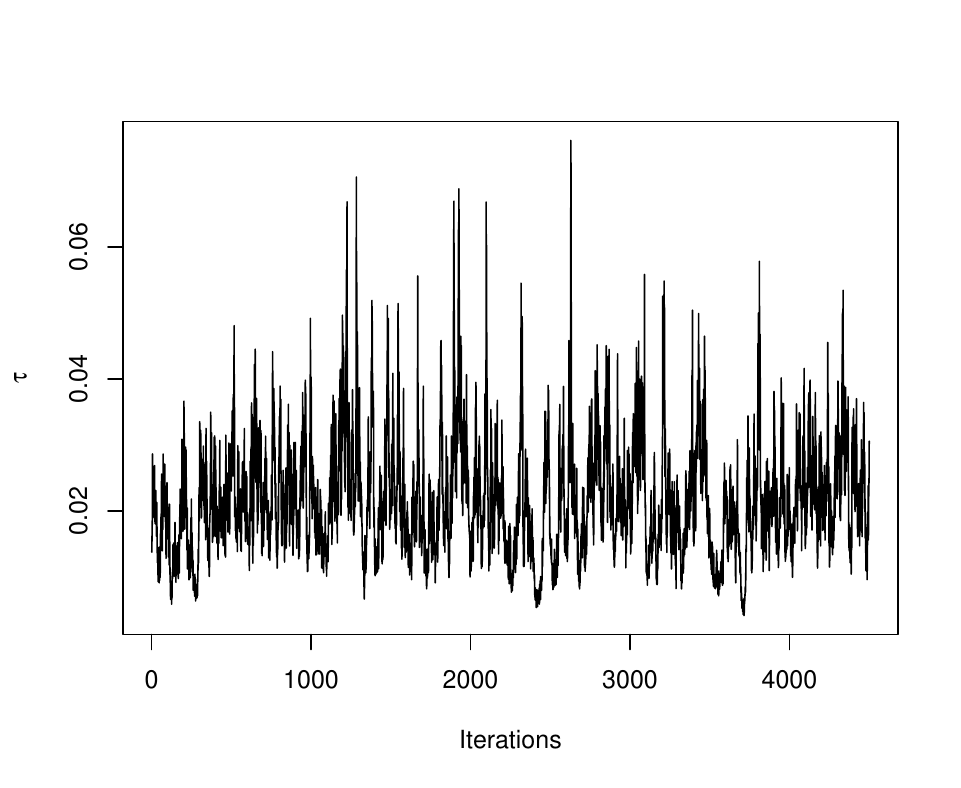}
				\caption{$\tau$}
				\label{fig:tau}
			\end{subfigure}
			\hfill
			\begin{subfigure}[b]{0.35\textwidth}
				\centering
				\includegraphics[width=\textwidth]{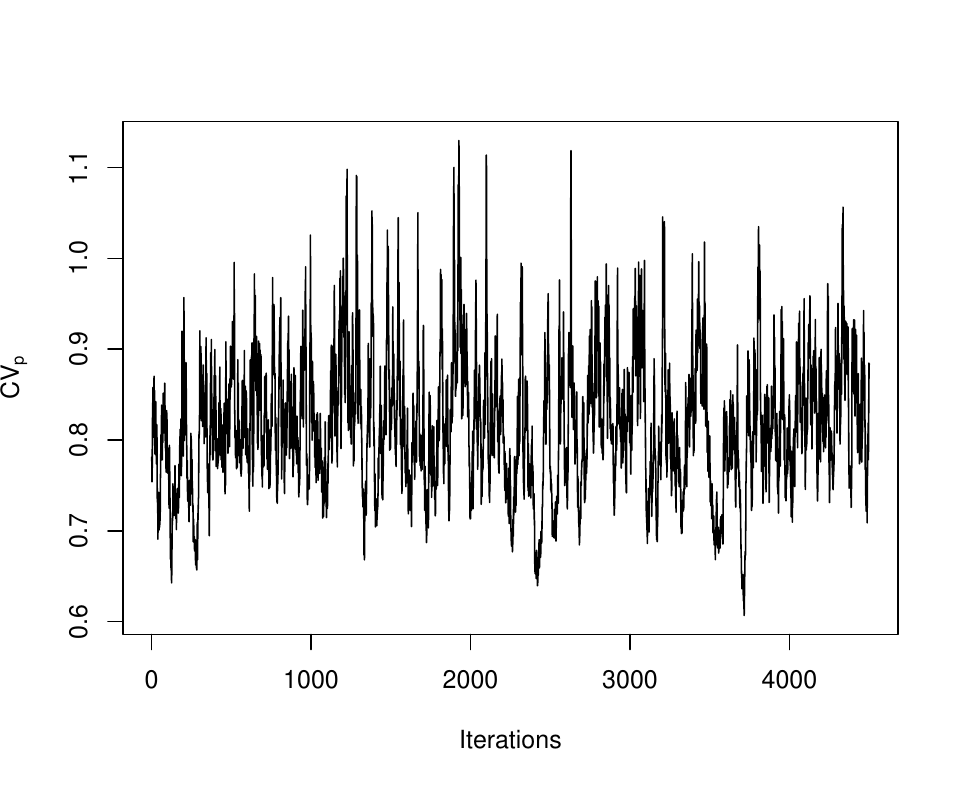}
				\caption{CV$_p$}
				\label{fig:CV}
			\end{subfigure}
			\hspace{1cm}
			\begin{subfigure}[b]{0.35\textwidth}
				\centering
				\includegraphics[width=\textwidth]{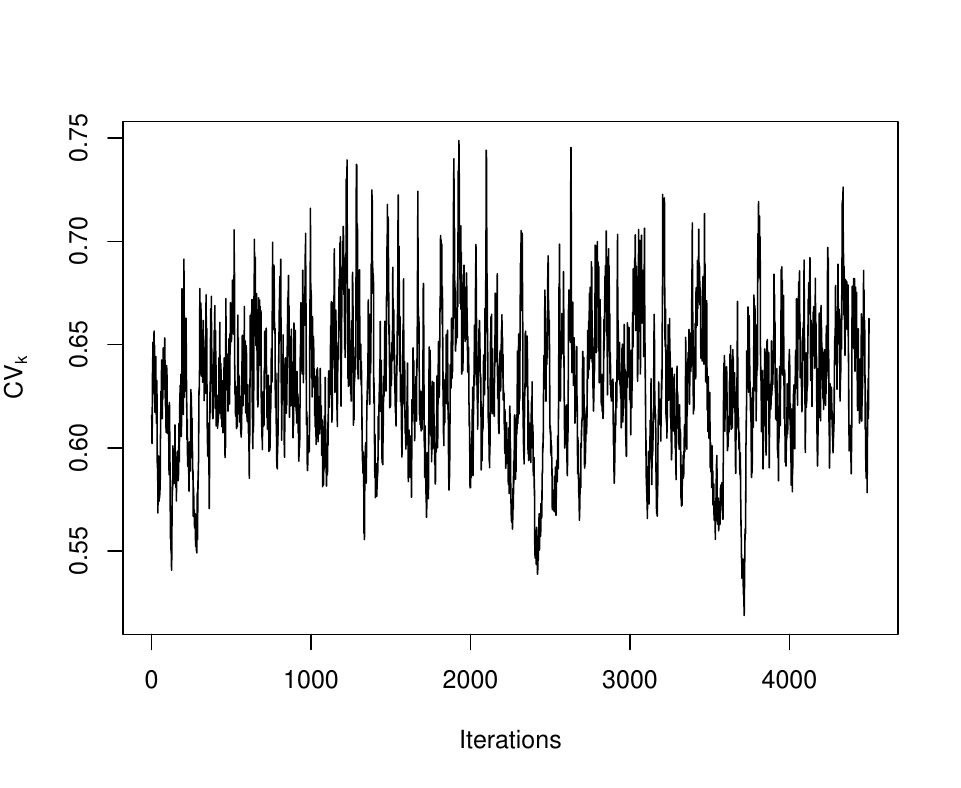}
				\caption{CV$_k$}
				\label{fig:kCV}
			\end{subfigure}
			\caption{Traceplots for the generated MCMC samples from each parameter's posterior distribution}
			\label{fig:traceplots}
		\end{figure}
		\begin{figure}[H]
			\centering
			\begin{subfigure}[b]{0.4\textwidth}
				\centering
				\includegraphics[width=\textwidth]{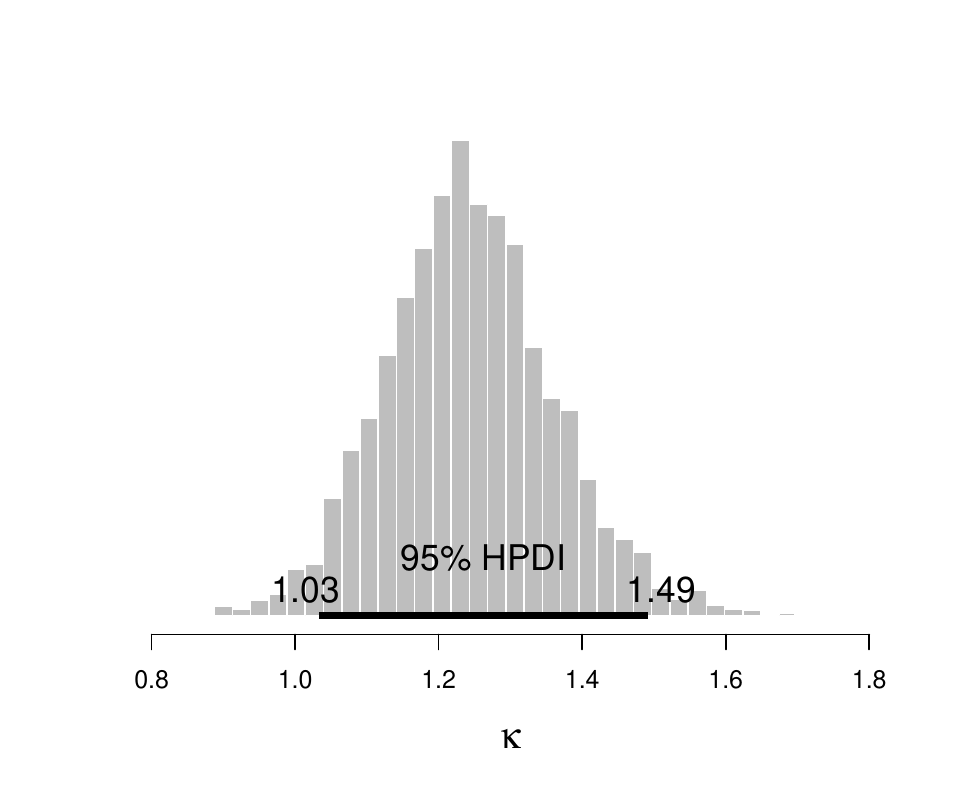}
				\caption{$\kappa$}
				\label{fig:hisK}
			\end{subfigure}
			\hspace{1cm}
			\begin{subfigure}[b]{0.4\textwidth}
				\centering
				\includegraphics[width=\textwidth]{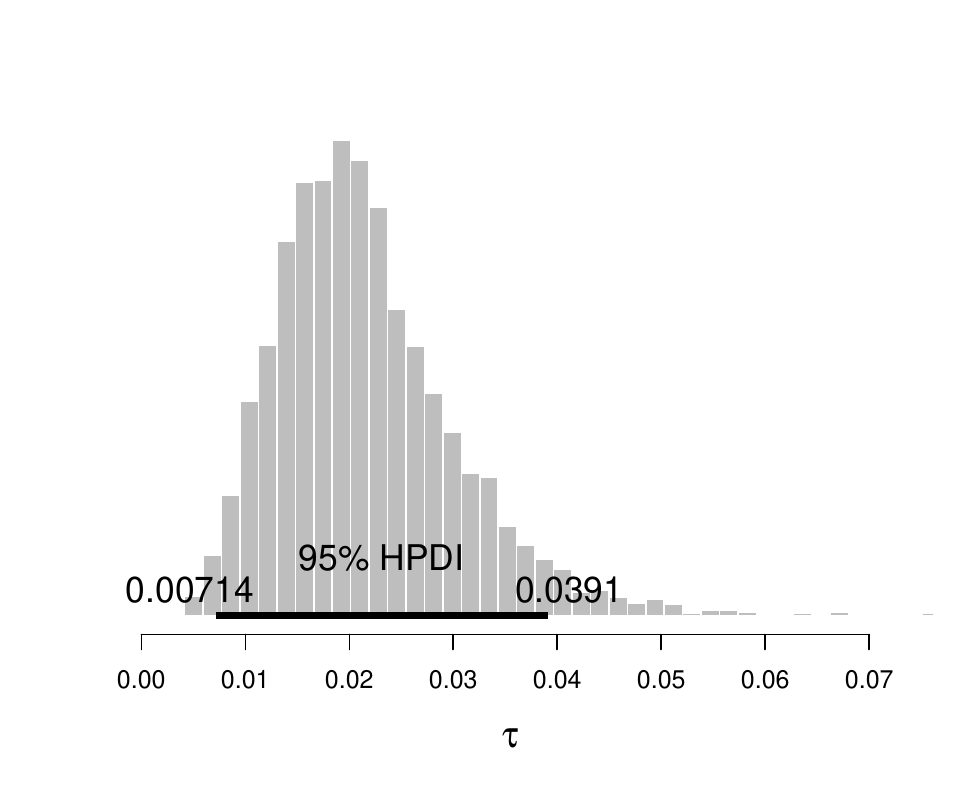}
				\caption{$\tau$}
				\label{fig:hisT}
			\end{subfigure}
			\hfill
			\begin{subfigure}[b]{0.4\textwidth}
				\centering
				\includegraphics[width=\textwidth]{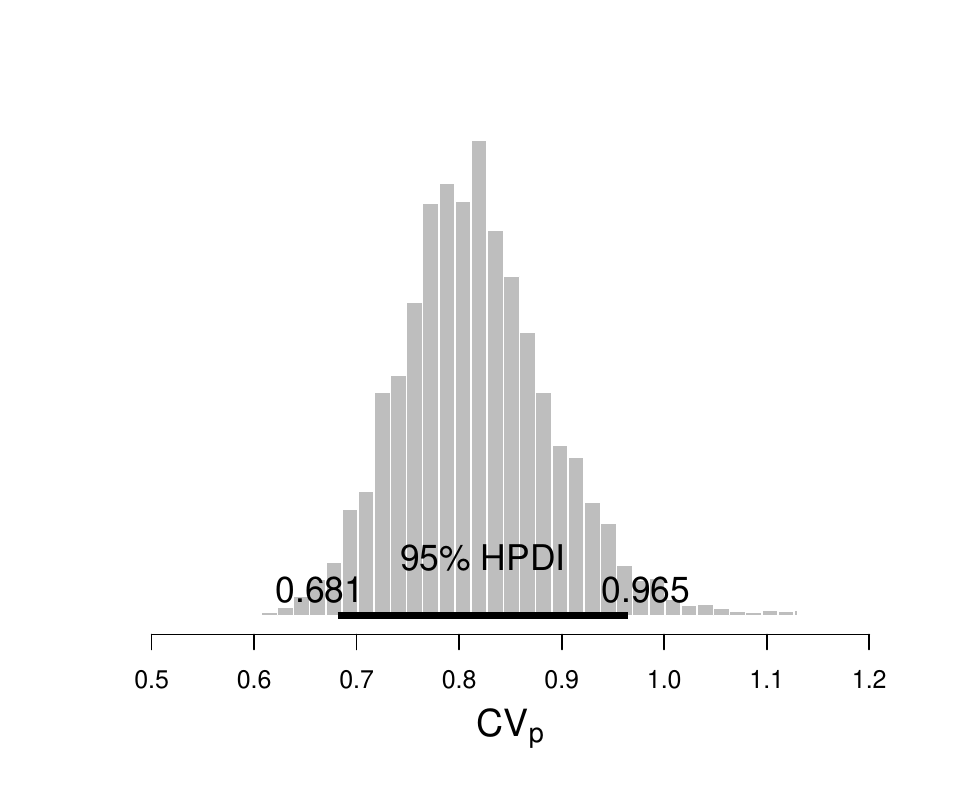}
				\caption{CV$_p$}
				\label{fig:hisCV}
			\end{subfigure}
			\hspace{1cm}
			\begin{subfigure}[b]{0.4\textwidth}
				\centering
				\includegraphics[width=\textwidth]{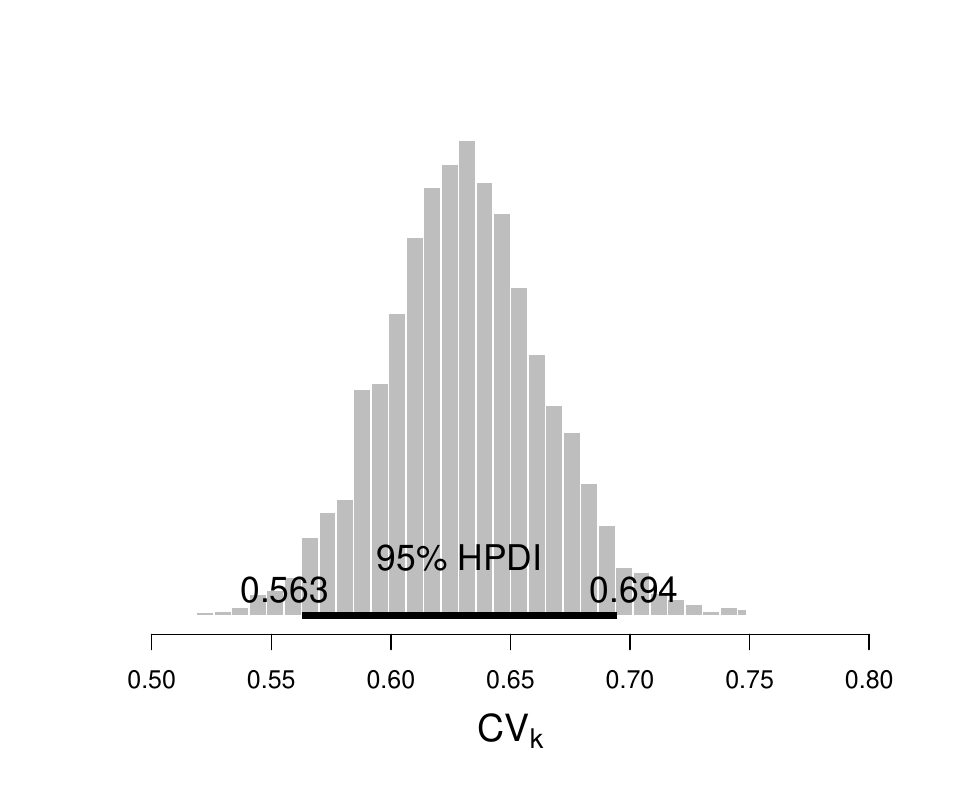}
				\caption{CV$_k$}
				\label{fig:hiskCV}
			\end{subfigure}
			\caption{Histograms representing the posterior distribution of the generated MCMC samples with the corresponding HPDIs}
			\label{fig:Histogram}
		\end{figure}
		\section{Conclusion}
		\indent In this article, we have discussed various methods for estimation of the coefficients of variation as well as the parameters of the Weibull distribution under the type-I progressively interval censoring scheme. Through the simulation study that we have conducted, we have observed that the Bayes estimators with informative gamma priors, and the proposed least squares estimators give the best performance for point estimation. Meanwhile, the HPDIs excel in giving best performance for interval estimation. Furthermore, the application of real data as an illustrative example supports the applicability of the proposed methods in practical scenarios. The proposed methods can also be applied to estimate the survival function and the hazard rate function. Future research could explore potential improvements and extensions. For instance, the approximate and generalized confidence intervals can be developed through the applicability of the least squares estimates. Furthermore, we hope to show that the better non-parametric estimators $\hat{F}$ may yield better estimators for the parameters in the least squares approximation. 
		\vspace{3mm} \\
		{\bf Conflict of interest}: The authors declare that they have no conflict of interest.
	
	\end{document}